%% file: main.tex
\documentclass[
  aps,             
  pra,              
  reprint, 
  superscriptaddress,% author affiliations as superscripts
  floatfix          
]{revtex4-2}
    
\usepackage{graphicx}   
\usepackage{amsmath,amssymb,amsfonts, amsthm}
\usepackage{mathtools}
\usepackage[hidelinks]{hyperref}
\usepackage{physics}
\usepackage{bm}
\usepackage{xcolor}
\usepackage{geometry}

\newcommand{\xx}[1]{}

\newcommand{\affchem}{Department of Chemistry, Purdue University, West Lafayette, Indiana 47907, USA}  
\newcommand{\affphys}{Department of Physics and Astronomy, Purdue University, West Lafayette, Indiana 47907, USA}
\newcommand{\affeng}{School of Electrical and Computer Engineering, Purdue University, West Lafayette, Indiana 47907, USA}
  
\begin{document}
\title{First order Maxwell operator formalism for macroscopic quantum electrodynamics}

\author{Ishita Agarwal}
\altaffiliation{These authors contributed equally to this work.}
\affiliation{\affphys}
\author{Ankit Kundu}
\altaffiliation{These authors contributed equally to this work.}
\affiliation{\affeng}
%\author{other authors}
\author{Christian M. Lange}
\author{Jonathan D. Hood}
\email{hoodjd@purdue.edu}
\affiliation{\affphys}
\affiliation{\affchem}
%\altaffiliation{$^*$Corresponding author: hoodjd@purdue.edu}

\date{\today}

\begin{abstract}
Standard macroscopic QED is built on the second-order Green's function for the electric field and discards open-system boundary terms. Here we develop a first-order electromagnetic operator approach that retains both $\mathbf{E}$ and $\mathbf{H}$ and keeps those boundary terms, naturally leading to a quantum input-output formalism. We recast Maxwell's equations as an operator equation for the dual field $\mathcal{E}=[\mathbf{E},\mathbf{H}]^T$, whose first-order Green operator $g$ propagates the electromagnetic state between surfaces. Symmetries of the Maxwell operator under energy and reciprocal inner products yield the propagation formula, Lorentz reciprocity, and a generalized optical theorem, with minimal vector calculus.
Quantizing via a Heisenberg--Langevin approach for absorptive, dispersive media yields two independent quantum noise sources: bulk Langevin operators from material absorption and input--output field operators at the boundary. Expressing the interior field in terms of these operators and the Green propagator yields an exact closed commutation relation $[\hat{\mathcal{E}},\hat{\mathcal{E}}^\dagger]\propto \mathrm{Im}\,g$, consistent with the fluctuation--dissipation theorem. This identity holds even when dielectrics extend to the boundary, as in waveguide input--output problems, and enables quantum input--output descriptions of complex photonic structures where the Green's function is obtained numerically, extending the framework beyond cavities and waveguides.
\end{abstract}

\maketitle

\section{Introduction}

%%%%%%%%%%%%% FIRST PARAGRAPH: BACKGROUND %%%%%%%%%%%%%%%%%%
While quantum optics began with the discrete modes of cavity QED~\cite{lodahl2015interfacing, glauber1963quantum, scully1997quantum, kimble1998strong, raimond2001manipulating, lange2026hybrid}, it has gradually moved toward more open systems. Waveguide QED introduced a continuum of propagating modes~\cite{shen2005coherent, martin-cano2011entanglement, sheremet2023waveguide}, and macroscopic QED, built on the electromagnetic Green's function, extended the theory to lossy dielectric environments and free-space geometries such as neutral atom arrays~\cite{asenjo-garcia2017atom, asenjo-garcia2017exponential, robicheaux2021beyond, masson2022universality, yelin2022superradiance, lange2024superradiant, Robicheaux2020directional, masson2024multilevelatom}. A growing class of nanophotonic structures used to couple to quantum emitters, for example photonic crystals, inverse-designed devices~\cite{molesky2018inverse}, and photonic bandgaps, do not fit naturally into modal descriptions but are more effectively captured by the electromagnetic Green's function, which is already the tool used to numerically design and optimize these devices.

% Experimental quantum optics has progressed beyond the regime of single emitters coupled to single optical modes to many-body dynamics interacting with structured light. Cavity QED established the interaction of single emitters with discrete photonic modes~\cite{lodahl2015interfacing, glauber1963quantum,scully1997quantum,kimble1998strong,raimond2001manipulating,lange2026hybrid}, and waveguide QED extended this to open, propagating channels~\cite{shen2005coherent,martin-cano2011entanglement,sheremet2023waveguide}. More recently, ordered neutral-atoms have demonstrated collective phenomena in free space and near photonic structures, and have exhibited a rich dependence on the geometry of the atomic arrays~\cite{asenjo-garcia2017atom,asenjo-garcia2017exponential,robicheaux2021beyond,masson2022universality,yelin2022superradiance,lange2024superradiant, Robicheaux2020directional, masson2024multilevelatom}. Similarly, advances in photonic inverse design and patterned metasurfaces have made it routine to couple quantum emitters to complex structured light environments that exceed simple input-output models~\cite{molesky2018inverse,solntsev2021metasurfaces}. 

\begin{figure}[tb!]
    \centering
    \includegraphics[width=1.\linewidth]{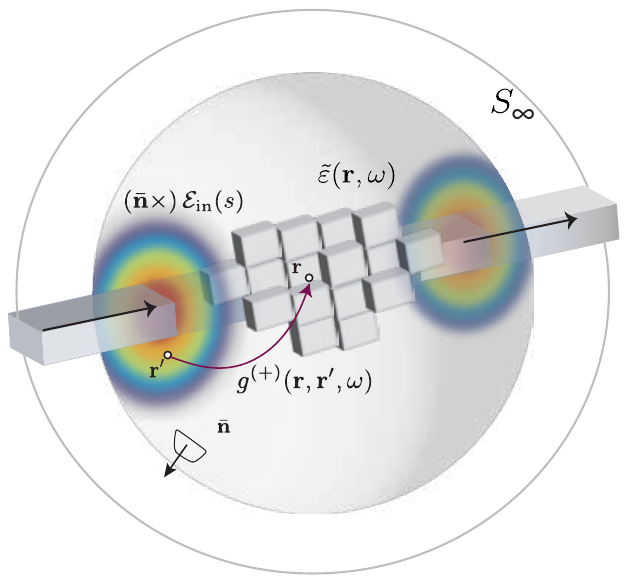}
    \caption{\textbf{Field decomposition geometry and system consideration}. A lossy, dispersive medium with material tensor $\bar{\varepsilon}(\mathbf{r},\omega)$ occupies a region $V$ enclosed by a surface $S$. The retarded Green operator $g^{(+)}(\mathbf{r},\mathbf{r}',\omega)$ propagates between interior points, while the incoming boundary field $(\bar{n}\times)\mathcal{E}_\mathrm{in}(s)$ enters through $S$. The outgoing radiation fields satisfy the Sommerfeld condition on $S_\infty$, rendering unique solutions.}
    \label{fig:field_geometry}
\end{figure}

%%%%%%%%%%%%% SECOND PARAGRAPH: THE PROBLEM %%%%%%%%%%%%%%%%%%
However, standard macroscopic QED~\cite{huttner1992quantization,gruner1996green-function,scheel2006quantum} is formulated in terms of noise operators associated with material absorption and effectively assumes that the field does not interact with a finite boundary. This approximation is justified when one considers a uniformly lossy medium filling all space. In that limit, the theory remains highly successful for local observables such as spontaneous emission rates and Casimir forces~\cite{buhmann2007dispersion}, but it does not explicitly include modes incident from the boundary or the corresponding boundary-induced fluctuations.
It was later shown that boundary operators are necessary in open scattering geometries to satisfy the fluctuation–dissipation theorem~\cite{drezet2017quantizing,ciattoni20240719quantum,stefano2001mode}, and moreover provide a natural route for introducing input light into the system through the boundaries. These modified Langevin noise (MLN) approaches demonstrated this for the case of free space extending to the boundary, expanding the incoming fields in plane waves. 
% However, nanophotonic input–output systems naturally have dielectrics extending to infinity, representing the input and output waveguides, where no plane-wave basis is available. 
By contrast, nanophotonic input-output platforms are typically coupled to dielectric waveguides or other structured photonic channels (Fig. \ref{fig:field_geometry}), for which a simple free-space plane-wave expansion is no longer available.

% In general, the theoretical framework underlying the description of complex light-matter interactions is based on the second-order Helmholtz equation for the electric field, with the dyadic Green's function as the central propagator~\cite{tai1994dyadic1, novotny2006principles}.
% This formalism has been enormously productive for local observables like spontaneous emission rates, local density of states, Casimir and dispersion forces~\cite{buhmann2007dispersion}. However, it carries $\mathbf{E}(\mathbf r)$ alone, and reconstructing the magnetic field needed for surface-to-surface propagation requires cumbersome vector calculus manipulations.
% On the quantum side, the standard macroscopic QED constructions~\cite{huttner1992quantization,gruner1996green-function,scheel2006quantum} were originally formulated for uniform lossy media that fill all space, sidestepping boundary contributions entirely.
% Extensions to finite objects~\cite{drezet2017quantizing,ciattoni20240719quantum,stefano2001mode} showed that boundary-assisted fluctuations are essential but relied on free-space plane-wave asymptotics at the enclosing surface, an assumption that does not hold for input--output problems in nanophotonic systems, where structured dielectric media extend to the far field regime. 

% %%%%%%%%%%%%% 3RD PARAGRAPH: THE SOLUTION %%%%%%%%%%%%%%%%%%
In this paper, we develop a Maxwell operator-based approach to macroscopic QED built on the first-order Maxwell equations~\cite{hanson2002operator,berreman1972optics,knoll1987action,silveirinha2021modal}. Whereas the traditional approach uses the second-order Helmholtz equation~\cite{tai1994dyadic1,novotny2006principles} and propagates only $\mathbf{E}$, our formalism retains the dual field $\mathcal{E} = [\mathbf{E}, Z_0\mathbf{H}]^T$, which leads to a compact propagation expression that relates interior fields to both volumetric sources and boundary data through a first-order Green operator. In this operator formalism, the symmetries of the Maxwell operator under energy and reciprocal inner products yield the key classical identities: the interior field representation, Lorentz reciprocity, and a generalized optical theorem, with minimal recourse to vector calculus.

We quantize the theory using a Heisenberg--Langevin approach~\cite{rosa2010electromagnetic} rather than the Fano diagonalization route underlying the Huttner--Barnett formulation~\cite{huttner1992quantization,gruner1996green-function,scheel2006quantum}. Although these two approaches have been shown to be equivalent~\cite{drezet2017equivalence}, the Langevin formulation is more convenient for our purposes because it keeps the field, bulk absorption noise, and incoming boundary fluctuations as explicitly distinct channels from the outset, making it naturally suited to an input--output description.
% Unlike Fano diagonalization which combines photonic, material, and reservoir degrees of freedom into a single polaritonic operator, the Langevin route keeps the noise sources individually identifiable (bulk absorption noise and incoming boundary fluctuations), making the formalism naturally suited to input–output descriptions. 
The interior quantum field takes the form
\begin{equation}
\begin{split}
\hat{\mathcal{E}}(\mathbf{r}) &=  k_0\int g(\mathbf{r},\mathbf{r}')\,\hat{\mathcal{P}}_N(\mathbf{r}')\,dV' \\
&\quad - i\oint g(\mathbf{r},\mathbf{s})\,(\bar{n}\times)\hat{\mathcal{E}}_\mathrm{in}(\mathbf{s})\,dS,
\end{split}
\end{equation}
where the polarization noise $\hat{\mathcal{P}}_N$ from the material absorption is integrated over the volume, and the tangential fields $(\bar{n} \times) \hat{\mathcal{E}}_\mathrm{in}$ is integrated over the boundary surface (see Fig.~\ref{fig:field_geometry}). We show that the commutation relations of the interior field written in terms of these operators satisfy the fluctuation–dissipation theorem exactly.  This yields a compact expression for the commutation relations between electric and magnetic fields in terms of the first-order Green operator: 
\begin{equation}
\bigl[\hat{\mathcal{E}}_i(\mathbf{r},\omega),\,\hat{\mathcal{E}}_j^\dagger(\mathbf{r}',\omega')\bigr] = \frac{\hbar k_0}{\pi\varepsilon_0}\,\mathrm{Im}\,g_{ij}(\mathbf{r},\mathbf{r}',\omega)\,\delta(\omega\!-\!\omega'),
  \label{eq:commutator_intro}
\end{equation}
where $\mathrm{Im}\,g \equiv (g - g^\dagger)/(2i)$. This commutator is a central result of the paper, showing that the first-order Green operator naturally encodes the quantum fluctuations and commutation structure of the electromagnetic field.

The Maxwell operator formalism minimizes the labor typically associated with exact full-wave treatments and allows the Green's function, including those obtained from numerical solvers, to serve as the central propagator in a quantum input–output scheme. Retaining both $\mathbf{E}$ and $\mathbf{H}$ yields exact propagation expressions between planes, enabling cascaded quantum systems built from heterogeneous photonic components. The formalism applies to a wide range of structures beyond traditional quantum optics, such as inverse-designed devices, slow-light waveguides, and chiral media, all of which have Green's functions readily computed by standard electromagnetic solvers.

Although the present work restricts attention to block-diagonal $\bar{\varepsilon}$, the first-order framework extends naturally to bianisotropic media with magnetoelectric coupling, including chiral, magneto-optic, and topological photonic systems, where the mixing of $\mathbf{E}$ and $\mathbf{H}$ in the constitutive relations makes a first-order treatment not merely convenient but essential~\cite{silveirinha2018topological}. A modal input–output theory extending the first-order formalism to guided modes will follow in subsequent work.

%%%%%%%%%%%% PAPER OUTLINE %%%%%%%%%%%%%%%%%%%%%%%%%%%%%%%
The paper is organized as follows. Sec.~\ref{mainsec:Symplectic_Formalism} introduces the first-order Maxwell operator, its Green's operator, and the connection to the conventional second-order dyadic formalism, then develops the energy and reciprocal inner products and derives the main classical identities. Sec.~\ref{sec:transfer} establishes the exact laws of plane-to-plane propagation. Sec.~\ref{sec:quantization} quantizes the theory, derives the field commutation relations, and establishes the quantum input–output result in Sec.~\ref{sec:quantum_io}.

% Section~\ref{sec:first_order} introduces the first-order Maxwell operator, section~\ref{sec:green's-operator} the Green operator, and section~\ref{sec:second-order} its relation to the conventional dyadic formalism.
% Section~\ref{sec:inner-product} develops the two pairings and derives the main classical identities, including the generalized optical theorem and the Stratton--Chu representation.
% Section~\ref{sec:quantization} quantizes the theory, derives the field commutation relations, and establishes the quantum input--output relation.
% We conclude by discussing implications for quantum scattering and cascaded propagation in complex nanophotonic environments.

%%%%%%%%%%%%%%%%%%%%%%%%%%%%%%%%%%%
\section{First-order symplectic formalism}
\label{mainsec:Symplectic_Formalism}

\subsection{First-Order Maxwell Operators}
\label{sec:first_order}

We begin by recasting Maxwell's equations in a first-order form that treats the electric and magnetic fields as components of a single dual field vector \cite{silveirinha2021modal, hanson2002operator, dudley1994mathematical, yoshibuhmann2013dispersion, ThomasWeiss2019FirstOrderPert}.
We work in the frequency domain with time dependence $e^{-i\omega t}$. Maxwell's equations in a linear, dispersive medium characterized by the relative permittivity and permeability tensors $\boldsymbol{\varepsilon}(\mathbf{r},\omega)$ and $\boldsymbol{\mu}(\mathbf{r},\omega)$ read
\begin{align}
\nabla \times \mathbf{E}(\mathbf{r},\omega) &= i\omega\mu_0\boldsymbol{\mu}(\mathbf{r},\omega)\,\mathbf{H}(\mathbf{r},\omega) - \mathbf{J}_M(\mathbf{r},\omega), \label{eq:maxwell_curl_E}\\
\nabla \times \mathbf{H}(\mathbf{r},\omega) &= -i\omega\varepsilon_0\boldsymbol{\varepsilon}(\mathbf{r},\omega)\,\mathbf{E}(\mathbf{r},\omega) + \mathbf{J}_E(\mathbf{r},\omega). \label{eq:maxwell_curl_H}
\end{align}
Here $\boldsymbol{\varepsilon}(\mathbf{r},\omega)$ and $\boldsymbol{\mu}(\mathbf{r},\omega)$ are dimensionless $3\times 3$ relative material tensors that are position-dependent, dispersive, and absorptive, with the vacuum response factored out through $\varepsilon_0$ and $\mu_0$. The frequency-domain fields are related to their time-domain counterparts by the Fourier transform $\mathbf{E}(\mathbf{r},\omega) = \int_{-\infty}^{\infty} \mathbf{E}(\mathbf{r},t)\,e^{i\omega t}\,dt$ and its inverse $\mathbf{E}(\mathbf{r},t) = \int_{-\infty}^{\infty} \mathbf{E}(\mathbf{r},\omega)\,e^{-i\omega t}\,d\omega/(2\pi)$, and similarly for all other fields and sources.
The electric current density $\mathbf{J}_E(\mathbf r, \omega)$ represents physical sources, i.e., displacement currents, while the magnetic current density $\mathbf{J}_M(\mathbf r, \omega)$ is fictitious but restores the duality between the two curl equations and will serve as the natural vehicle for representing equivalent surface sources.

We combine the fields and sources into dual vectors
\begin{equation}
\mathcal{E}(\mathbf{r},\omega) \equiv \begin{bmatrix} \mathbf{E}(\mathbf{r},\omega) \\ Z_0\mathbf{H}(\mathbf{r},\omega) \end{bmatrix}, \,\, \mathcal{J}(\mathbf{r},\omega) \equiv \begin{bmatrix} Z_0\mathbf{J}_E(\mathbf{r},\omega) \\ \mathbf{J}_M(\mathbf{r},\omega) \end{bmatrix}.
\label{eq:dual_field_def}
\end{equation}
We normalize the magnetic field by $Z_0=\sqrt{\mu_0/\varepsilon_0}$ so that both components of the dual field carry the same dimensions.
The antisymmetric coupling between $\mathbf{E}$ and $\mathbf{H}$ in Maxwell's equations is captured by defining the dual curl operator
\begin{equation}
\bar{\nabla}\times \;\equiv\; J \otimes (\nabla\times) = \begin{bmatrix} 0 & \nabla\times \\ -\nabla\times & 0 \end{bmatrix},
\label{eq:dual_curl_def}
\end{equation}
where
\begin{equation}
J \equiv \begin{bmatrix} 0 & 1 \\ -1 & 0 \end{bmatrix}
\label{eq:symplectic_matrix}
\end{equation}
is the $2\times 2$ symplectic matrix, satisfying $J^2 = -\mathbf{1}$ and $J^\dagger = -J$. We also define the dimensionless dual relative material tensor

\begin{equation}
\bar{\varepsilon}(\mathbf{r},\omega) \equiv \begin{bmatrix} \boldsymbol{\varepsilon}(\mathbf{r},\omega) & 0 \\ 0 & \boldsymbol{\mu}(\mathbf{r},\omega) \end{bmatrix}.
\label{eq:dual_material_def}
\end{equation}

With these definitions, the two Maxwell curl equations combine into a single first-order operator equation
\begin{equation}
\Big(\bar{\nabla}\times + i k_0\,\bar{\varepsilon}(\mathbf{r},\omega)\Big)\,\mathcal{E}(\mathbf{r},\omega) = \mathcal{J}(\mathbf{r},\omega).
\label{eq:maxwell_combined}
\end{equation}
Here $k_o =\omega/c$ is the free-space wave-vector. Multiplying throughout by $i$, we define a dual curl operator
\begin{equation}
\mathcal{H} \equiv i\,\bar{\nabla}\times 
\label{eq:maxwell_hamiltonian}
\end{equation}
so that Maxwell's equations become
\begin{equation}
\Big(\mathcal{H} - k_0\,\bar{\varepsilon}(\mathbf{r},\omega)\Big)\,\mathcal{E}(\mathbf{r},\omega) = i\,\mathcal{J}(\mathbf{r},\omega).
\label{eq:maxwell_hamiltonian_form}
\end{equation}
The factor of $i$ is chosen so that $\mathcal{H}$ is formally self-adjoint under the  $L^2$ inner product, as will be established later.

We also define the Maxwell operator
\begin{equation}
\mathcal{M} \equiv \mathcal{H} - k_0\bar{\varepsilon}
\label{eq:maxwell_operator_def}
\end{equation}
and the source $\mathcal{S}(\mathbf r) \equiv i\,\mathcal{J}(\mathbf r),$
so that the first-order Maxwell system reads
\begin{equation}
\mathcal{M}\,\mathcal{E}(\mathbf r) = \mathcal{S}(\mathbf r)
\label{eq:maxwell_first_order}
\end{equation}

The operator $\mathcal{M}$ can be viewed as an operator that maps dual fields to dual sources; inverting it with appropriate boundary conditions will give us the Green operator, which propagates from the sources to the resulting fields.

% \paragraph{Unweighted equation.}
% This is a weighted eigenvalue-type equation in which the unbounded differential operator $\mathcal{H}$ is modified by the frequency- and position-dependent material weight $k_0\,\bar{\varepsilon}$. One can also define an unweighted eigenvalue problem by absorbing $\bar{\varepsilon}$ into a rescaled Hamiltonian $\tilde{\mathcal{H}} = \bar{\varepsilon}^{-1/2}\,\mathcal{H}\,\bar{\varepsilon}^{-1/2}$~\cite{sakoda2004optical, Flow_of_light_Joannopoulos}, which brings the system into the standard resolvent form $(\tilde{\mathcal{H}} - k_0)|\tilde{\mathcal{E}}\rangle = |\tilde{\mathcal{S}}\rangle$.

\paragraph{Second-order Maxwell operator.}
While we work primarily in first-order form, the second-order Helmholtz equation is recovered by left-multiplying $\mathcal{M}\mathcal{E} = \mathcal{S}$ by $(\mathcal{H}+k_0\bar{\varepsilon})\bar{\varepsilon}^{-1}$. The cross terms $\pm k_0\mathcal{H}$ cancel in the product, giving the second-order Maxwell operator
\begin{equation}
\begin{split}
\mathcal{M}^{(2)} &= (\mathcal{H}+k_0\bar{\varepsilon})\,\bar{\varepsilon}^{-1}\,(\mathcal{H}-k_0\bar{\varepsilon}) \\
&= \mathcal{H}\,\bar{\varepsilon}^{-1}\,\mathcal{H} - k_0^2\bar{\varepsilon}.
\end{split}
\end{equation}
The second-order equation is then
\begin{equation}
\mathcal{M}^{(2)}\,\mathcal{E} = \left(\mathcal{H} + k_0\bar{\varepsilon}\right)\bar{\varepsilon}^{-1}\,\mathcal{S}.
\label{eq:second_order_equation}
\end{equation}

Because $\mathcal{H}$ is off-diagonal and $\bar{\varepsilon}$ is block diagonal, the operator $\mathcal{M}^{(2)}$ is itself block diagonal~\cite{johnson2002adiabatic}, and the system therefore decouples into two independent vector Helmholtz equations,
\begin{align}
&\nabla\!\times\boldsymbol{\mu}(\mathbf r)^{-1}
  \nabla\!\times\mathbf{E}(\mathbf r)
  - k_0^2\boldsymbol{\varepsilon}\,\mathbf{E}(\mathbf r)
  \nonumber\\
&\quad= ik_0 Z_0\mathbf{J}_E(\mathbf r)
  - \nabla\!\times\boldsymbol{\mu}^{-1}
    \mathbf{J}_M(\mathbf r),
\\[4pt]
&\nabla\!\times\boldsymbol{\varepsilon}(\mathbf r)^{-1}
  \nabla\!\times(Z_0\mathbf{H}(\mathbf r))
  - k_0^2\boldsymbol{\mu}\,(Z_0\mathbf{H}(\mathbf r))
  \nonumber\\
&\quad= ik_0\mathbf{J}_M(\mathbf r)
  + \nabla\!\times\boldsymbol{\varepsilon}^{-1}
    (Z_0\mathbf{J}_E(\mathbf r)).
    \label{eq:second_order_maxwell}
\end{align}

These are the standard equations underlying the dyadic Green's function formalism used in classical electrodynamics and macroscopic QED~\cite{tai1994dyadic1,novotny2006principles}; the connection to the first-order Green's operator and a translation dictionary between the two formalisms are given in Sec.~\ref{sec:second-order}.

The passage from first to second order comes at a cost.  The source terms acquire curl derivatives of the original currents, the self-adjoint structure of the first-order operator is lost, and the decoupling of $\mathbf{E}$ and $\mathbf{H}$ makes the  plane-to-plane propagation less direct (Sec.~\ref{sec:transfer}).
%, a capability that is less direct in the $3\times 3$ dyadic framework, which must reconstruct the missing field component.
%Moreover the factorization, $(\mathcal{H}+k_0\bar{\varepsilon})\bar{\varepsilon}^{-1}(\mathcal{H}-k_0\bar{\varepsilon})$, is a same-frequency elimination identity, not a physical branch decomposition. Hence, the complementary factor $(\mathcal{H}+k_0\bar{\varepsilon}(\omega))$ differs from the true negative-frequency operator $\mathcal{M}(-\omega) = \mathcal{H}+k_0\bar{\varepsilon}^*(\omega)$ in any dispersive medium, so $\mathcal{M}^{(2)} \neq \mathcal{M}(-\omega)\mathcal{M}(\omega)$.
The first-order Maxwell operator retains the branch structure, directional information, and symplectic pairing, making it particularly well suited to propagation, reciprocity, and input–output theory.

%%%%%%%%%%%%%%%%%%%%%%%%%%%%%%%%%%%
\subsection{Green's Operator and Its Kernel}
\label{sec:green's-operator}
\paragraph{Operator representation.}
We adopt a Dirac-notation convention in which the dual field is
promoted to an abstract ket $|\mathcal{E}\rangle$, with its
position-space projection defined by
\begin{equation}
\mathcal{E}(\mathbf{r},\omega) \equiv \langle\mathbf{r}|\mathcal{E}\rangle,
\qquad
\mathcal{S}(\mathbf{r},\omega) \equiv \langle\mathbf{r}|\mathcal{S}\rangle.
\end{equation}
This is in direct analogy with quantum mechanics, where one writes
the abstract state $|\psi\rangle$ and recovers the wavefunction only
upon projecting onto position eigenstates,
$\psi(\mathbf{r}) = \langle\mathbf{r}|\psi\rangle$.  The ket $|\mathcal{E}\rangle$ is not a normalizable Hilbert-space vector but rather a generalized state on which $\mathcal{M}$ acts in a weak sense, well defined when sources are locally square-integrable and the appropriate radiation condition is imposed at infinity.

Just as the momentum operator acts abstractly as $\hat{p}|\psi\rangle$ while its position-space
projection yields the differential operator
$\langle\mathbf{r}|\hat{p}|\psi\rangle = -i\hbar\nabla\psi(\mathbf{r})$,
the Maxwell operator $\mathcal{M}$ acts abstractly on kets while its position-space projection reads
\begin{equation}
\langle\mathbf{r}|\mathcal{M}|\mathcal{E}\rangle
= \bigl(i\,\bar{\nabla}\times
  - k_0\,\bar{\varepsilon}(\mathbf{r},\omega)\bigr)
  \mathcal{E}(\mathbf{r},\omega).
\end{equation}
With this notation, Maxwell's equations take the operator form $\mathcal{M}|\mathcal{E}\rangle = |\mathcal{S}\rangle$.

To solve this for arbitrary sources $|\mathcal{S}\rangle$ we seek a Green operator $G$ such that $|\mathcal{E}\rangle = G|\mathcal{S}\rangle$. However, the inverse of $\mathcal{M}$ is not unique~\cite{hanson2002operator}, the homogeneous equation $\mathcal{M}|\mathcal{E}\rangle=0$ admits nontrivial solutions, the radiation modes of free space, so one must specify boundary conditions to single out a unique response~\cite{chew1995waves1, collin1990field}.
The two standard choices are the outgoing (retarded) boundary condition, in which radiation emitted by the source travels outward to infinity, and the incoming (advanced) boundary condition, in which radiation arrives from infinity and propagates toward the source.

A systematic way to enforce either choice is the limiting absorption principle, in which an infinitesimal uniform loss or gain is added to the material response:
\begin{equation}
  \mathcal{M}^{(\pm)} \equiv \mathcal{H} - k_0\bar{\varepsilon} \mp i\eta,
  \qquad \eta\to 0^+.
\end{equation}
The $-i\eta$ term introduces infinitesimal uniform loss, under which all fields decay exponentially at large distances and only outgoing solutions remain square-integrable; $+i\eta$ selects the incoming solution by the same argument, but with uniform gain.
The outgoing (retarded, $+$) and incoming (advanced, $-$) Green's operators are then the inverse of the Maxwell operator under the limiting absorption principle,
\begin{equation}
  G^{(\pm)} = \lim_{\eta\to 0^+}
  \bigl(\mathcal{H} - k_0\bar{\varepsilon} \mp i\eta\bigr)^{-1}.
\end{equation}
The operator $G^{(\pm)}$ maps a source $\mathcal{S}$ to the corresponding field $|\mathcal{E}\rangle=G^{(\pm)}|\mathcal{S}\rangle$ and satisfies 
\begin{equation}
    \mathcal M^{\pm}G^{\pm}=\mathbb{I}
\end{equation}
In a typical scattering problem, a compactly supported source is square-integrable but the radiated field it produces is not, since the field carries energy to infinity. For finite $\eta$ the exponential damping renders both the source and the field square-integrable, so the inverse is well defined.
%in the limit $\eta\to 0^+$ this regularization is removed.  The resulting fields are locally but not globally square-integrable; the precise functional-space setting is deferred to Sec.~\ref{sec:inner-product}.

\paragraph{Kernel representation.}
In order to transfer from the operator equation $|\mathcal{E}\rangle=G^{(+)}|\mathcal{S}\rangle$ to the position space, we project both sides onto the position--component basis.
Let $|\mathbf{r},i\rangle$ denote the basis ket at position $\mathbf{r}$ with dual-field index $i=1,\dots,6$, where $i$ runs over the three Cartesian components of $\mathbf{E}$ and of $Z_0\mathbf{H}$. These kets satisfy the completeness relation
\begin{equation}
\sum_{i}\int dV\,|\mathbf{r},i\rangle\langle\mathbf{r},i|=\mathbb{I}.
\end{equation}
Projecting the left-hand side gives $\langle\mathbf{r},i|\mathcal{E}\rangle = \mathcal{E}_i(\mathbf{r})$, which recovers the position-space field introduced in Sec.~\ref{sec:first_order}. On the right-hand side, inserting the resolution of the identity between $G^{(+)}$ and $|\mathcal{S}\rangle$ gives
\begin{equation}
  \mathcal{E}_i(\mathbf{r}, \omega)
  = \sum_j\int dV'\;
    \langle\mathbf{r},i|G^{(+)}|\mathbf{r}',j\rangle\;
    \mathcal{S}_j(\mathbf{r}',\omega),
\end{equation}
where we identify the Green's kernel $g^{(+)}_{ij}(\mathbf{r},\mathbf{r}',\omega) \equiv \langle\mathbf{r},i|G^{(+)}|\mathbf{r}',j\rangle$.

In position space, the operator $G^{(+)}$ acts through the $6\times 6$ kernel $g^{(+)}(\mathbf{r},\mathbf{r}',\omega)$:
\begin{equation}
  \mathcal{E}(\mathbf{r})
  = \int g^{(+)}(\mathbf{r},\mathbf{r}',\omega)\,
    \mathcal{S}(\mathbf{r}',\omega)\,dV'.
\end{equation}
The kernel satisfies
\begin{equation}
    \mathcal{M}^{(\pm)}_{\mathbf{r}}\,g^{(\pm)}(\mathbf{r},\mathbf{r}',\omega) = \mathbb{I}_6\,\delta^{(3)}(\mathbf{r}-\mathbf{r}')
\end{equation}
Here, the subscript on $\mathcal{M}^{(\pm)}_{\mathbf{r}}$ indicates that the differential operator acts on the variable $\mathbf{r}$ as opposed to $\mathbf{r^\prime}$.
%This correspondence between operators and kernels, mediated by the resolution of the identity, will recur when we construct the adjoint of $g$ and derive the optical theorem.

In the dual field formulation, the kernel has a $6\times 6$ block structure,
\begin{equation}
  g^{(\pm)}(\mathbf{r},\mathbf{r}',\omega)
  = \begin{bmatrix}
      g_{EJ} & g_{EM} \\[2pt] g_{HJ} & g_{HM}
    \end{bmatrix}(\mathbf{r},\mathbf{r}',\omega),
\end{equation}
which couples each output field ($\mathbf{E}$ or $\mathbf{H}$) to each source type (electric $\mathbf{J}$ or magnetic $\mathbf{M}$), with off-diagonal blocks absent in the conventional second-order dyadic Green's function~\cite{collin1986dyadic}.

% \paragraph{Propagation from fields at boundaries}
%\paragraph{Propagation from boundary fields}
The Green's operator also propagates the fields between surfaces. 
Love's equivalence principle states that the tangential field on a boundary may be interpreted as an equivalent surface current encoded by 
\begin{equation}
\mathcal{J}(s)
= (\bar{n}\times)\mathcal{E}(\mathbf{s})
\equiv
\begin{pmatrix}
  \hat{\mathbf{n}}\times Z_0\mathbf{H}(\mathbf{s})\\
  -\hat{\mathbf{n}}\times\mathbf{E}(\mathbf{s})
\end{pmatrix},
\quad \mathbf{s}\in S.
\label{eq:tangential_trace}
\end{equation}
This is the tangential dual trace  containing both the electric and magnetic surface-current contributions. The retarded Green operator then propagates this surface source into the interior, just as it propagates bulk sources distributed in the volume. This is shown more explicitly in Sec.~\ref{sec:stratton_chu_l2}.  The final result is that the interior field can be written as~\cite{schelkunoff1936some,collin1990field}
\begin{align}
\mathcal{E}(\mathbf{r})
= \int_V &g^{(+)}(\mathbf{r},\mathbf{r}')\,\mathcal{S}(\mathbf{r}')\,dV'
   \nonumber\\
& -i\oint_S g^{(+)}(\mathbf{r},\mathbf{s})\,
    (\bar{n}\times\mathcal{E})(\mathbf{s})\,dS,
\label{eq:stratton_chu_intro}
\end{align}
% which decomposes the field into a volume contribution from bulk sources $\mathcal S$ and a surface contribution from the equivalent Love currents on $S$~\cite{schelkunoff1936some,collin1990field}. 
where the prefactor in the second term is an artifact of the convention of the curl operator written with $i$ and $\bar{n}$ assumed to be the dual outward normal to the surface defined in subsequent sections. 

Although the boundary field appearing in Eq.~\eqref{eq:stratton_chu_intro} is formally the total field, it can in general be decomposed into ingoing and outgoing contributions. For an exterior boundary, the outgoing component does not influence the interior and may therefore be neglected. This separation becomes especially clear in a modal decomposition of the field, for example in a lossless waveguide.

For interior surfaces, however, the same equation admits a different and equally useful reading: rather than projecting onto the incident part alone, one may regard it as an exact propagation law for the total field between surfaces. In this form, the first-order Green operator propagates the full tangential electromagnetic field, naturally leading to a transfer-matrix description of the total field.
\subsection{Second-Order Green's Function and Relation to the First Order}
\label{sec:second-order}

The conventional approach to macroscopic electrodynamics works with the second-order Helmholtz equation and its associated $3\times 3$ dyadic Green's functions. The first-order formalism does not require these objects, but we derive the connection to provide a translation dictionary between the two frameworks.

The second-order Maxwell operator $\mathcal{M}^{(2)} = \mathcal{H}\bar{\varepsilon}^{-1}\mathcal{H} - k_0^2\bar{\varepsilon}$, derived in Eq.~\eqref{eq:second_order_maxwell}, is block diagonal, so its Green's operator, defined by
\begin{equation}
  \mathcal{M}^{(2)} G^{(2,\pm)} = \mathbb{I},
\end{equation}
with the same limiting absorption prescription as before, is likewise block diagonal.
The corresponding kernel satisfies
\begin{equation}  \mathcal{M}^{(2)}_{\mathbf{r}}\,g^{(2,\pm)}(\mathbf{r},\mathbf{r}',\omega)
  = \mathbb{I}_6\,\delta^{(3)}(\mathbf{r}-\mathbf{r}'),
\end{equation}
and has the block structure
\begin{equation}
  g^{(2,\pm)}(\mathbf{r},\mathbf{r}',\omega)
  = \begin{bmatrix}
      g_E^{(\pm)}(\mathbf{r},\mathbf{r}',\omega) & 0 \\
      0 & g_H^{(\pm)}(\mathbf{r},\mathbf{r}',\omega)
    \end{bmatrix},
\end{equation}
where $g_E$ and $g_H$ are the $3\times 3$ dyadic Green's functions of the electric and magnetic vector Helmholtz equations, respectively.

To connect the two formalisms, we note that applying $(\mathcal{H}+k_0\bar{\varepsilon})\bar{\varepsilon}^{-1}$ to the first-order equation $\mathcal{M}|\mathcal{E}\rangle=|\mathcal{S}\rangle$ yields 
\begin{equation}
    \mathcal{M}^{(2)}|\mathcal{E}\rangle = (\mathcal{H}+k_0\bar{\varepsilon})\bar{\varepsilon}^{-1}|\mathcal{S}\rangle.
\end{equation}
Comparing $|\mathcal{E}\rangle = G^{(+)}|\mathcal{S}\rangle$ and $|\mathcal{E}\rangle = G^{(2,+)}(\mathcal{H}+k_0\bar{\varepsilon}) \bar{\varepsilon}^{-1}|\mathcal{S}\rangle$ for arbitrary $\mathcal{S}$, the operator identity follows:
\begin{equation}
  G^{(\pm)} = G^{(2,\pm)}
  \bigl(\mathcal{H}+k_0\bar{\varepsilon}\bigr)
  \bar{\varepsilon}^{-1}.
  \label{eq:first_second_relation}
\end{equation}
Passing to position representation and integrating by parts to move $\mathcal{H}_{\mathbf{r}'}$ onto the kernel (the surface term vanishes for compactly supported sources) gives the block-component relations
\begin{align}
  g_{EJ}^{(\pm)}(\mathbf{r},\mathbf{r}',\omega)
    &= k_0\, g_{E}^{(\pm)}(\mathbf{r},\mathbf{r}',\omega),
    \label{eq:gEJ}\\
  g_{EM}^{(\pm)}(\mathbf{r},\mathbf{r}',\omega)
    &= i\bigl[\nabla_{\mathbf{r}'}\times
       g_{H}^{(\pm)}(\mathbf{r},\mathbf{r}',\omega)\bigr]\,
       \boldsymbol{\mu}^{-1}(\mathbf{r}'),
    \label{eq:gEM}\\
  g_{HJ}^{(\pm)}(\mathbf{r},\mathbf{r}',\omega)
    &= -i\bigl[\nabla_{\mathbf{r}'}\times
        g_{E}^{(\pm)}(\mathbf{r},\mathbf{r}',\omega)\bigr]\,
        \boldsymbol{\varepsilon}^{-1}(\mathbf{r}'),
    \label{eq:gHJ}\\
  g_{HM}^{(\pm)}(\mathbf{r},\mathbf{r}',\omega)
    &= k_0\, g_{H}^{(\pm)}(\mathbf{r},\mathbf{r}',\omega),
    \label{eq:gHM}
\end{align}
where the curl acts on the kernel before contraction with the inverse relative material tensor.
The diagonal blocks are $k_0$ times the standard dyadic Green's functions, while the off-diagonal blocks, absent in the second-order formalism, encode the cross-coupling between field type and source type through a single curl and the inverse relative material tensor.

These relations provide a direct dictionary between the two formalisms, allowing results from the extensive second-order literature to be translated into the first-order framework and vice versa.
%The second-order dyadic framework is not incomplete in principle: it encodes the same physical solutions and remains entirely adequate for many standard observables. The distinction is one of structure and convenience.

% In the first-order representation the full electromagnetic state, its adjoint properties, and its boundary traces are all carried explicitly, whereas in the second-order formulation these features are distributed across separate objects and must be reconstructed when needed. 
% This difference becomes especially significant when one wishes to formulate exact surface-to-surface propagation laws or quantum input--output relations, where the full tangential Cauchy data $(\hat{\mathbf{n}}\times Z_0\mathbf{H},\,-\hat{\mathbf{n}}\times\mathbf{E})$ are the relevant boundary variables.

%%%%%%%%%%%%%%%%%%%%%%%%%%%%%%%%%%%%%%%%%%%%%
\subsection{Introduction to adjoint identities}
\label{sec:inner-product}
%%%%%%%%%%%%%%%%%%%%%%%%%%%%%%%%%%%%%%%%%%%%%

In electrodynamics, inner products are conventionally used to calculate energy or, in lossless systems, to find eigenmodes with real eigenvalues. When the Maxwell operator is self-adjoint under a given inner product, eigenmodes are orthogonal and eigenvalues are real. In this paper we pursue a different goal: we use inner products to uncover relations between the interior field and its values at the boundary.  

The procedure is as follows. For any operator $\mathcal{O}$ acting within an inner product space, the adjoint $\mathcal{O}^\dagger$ is defined by
\begin{equation} \label{eq:adjoint}
\langle\mathcal{E}_1|\mathcal{O}\mathcal{E}_2\rangle = \langle\mathcal{O}^\dagger\mathcal{E}_1|\mathcal{E}_2\rangle + \mathcal{B}[\mathcal{E}_1,\mathcal{E}_2].
\end{equation}
When $\mathcal{O}$ is a differential operator, moving it from one slot to the other requires integration by parts, producing the boundary term $\mathcal{B}$.  
%This is analog  of the 1D case $\int u\,\partial_x v\,dx = -\int (\partial_x u)\,v,dx + [uv]{\partial\Omega}$. 
In closed systems this boundary term is typically set to zero by imposing vanishing-field conditions at the boundary. In open systems, however, the fields do not vanish at the boundary, and $\mathcal{B}$ encodes a physically meaningful relation between the interior fields and their boundary values.
% It is precisely this term that will give rise to the surface contributions in the optical theorem, the propagation equation, and ultimately the quantum input–output channels.

When the operator is self-adjoint under a particular inner product, corresponding to a symmetry of the system, Eq.~\ref{eq:adjoint} then becomes a relation between the field in the interior and at the boundary. These are the operator analogs of the vector Green's identities, derived here from symmetries of the Maxwell operator rather than from vector calculus.

We introduce two inner products: the energy inner product (the standard $L^2$ norm) in Sec. \ref{sec:energy_adjoint} and the reciprocal inner product in Sec. \ref{sec:reciprocity}. Together they yield the propagation formula, the generalized optical theorem, and Lorentz reciprocity, all relating fields and sources in the volume to their values at the surface. These identities are the classical precursors that later allow us to close the commutation relations for $\hat{\mathcal{E}}$ and $\hat{\mathcal{E}}^\dagger$, showing that the total quantum fluctuations arise from two sources: absorption in the volume and fields at the boundary.

\subsection{Energy inner product and adjoint operators}
\label{sec:energy_adjoint}
%%%%%%%%%%%%%%%%%%%%%%%%%%%%%%%%%%%

The standard $L^2$ inner product of dual fields is
\begin{equation}
\begin{split}
\langle \mathcal{E}_1|&\mathcal{E}_2\rangle
= \int \mathcal{E}_1^\dagger(\mathbf{r})\,
   \mathcal{E}_2(\mathbf{r})\,dV\\
&= \int \bigl(\mathbf{E}_1^*(\mathbf r)\!\cdot\mathbf{E}_2(\mathbf r)
   + Z_0^2\,\mathbf{H}_1^*(\mathbf r)\!\cdot\mathbf{H}_2(\mathbf r)\bigr)\,dV.
\end{split}
\label{eq:L2ip}
\end{equation}
% This pairing is sesquilinear and positive definite. 
We note that this inner product is a mathematical tool for operator identities; it is not the physical electromagnetic energy in dispersive media~\cite{landau1984electrodynamics}. 
% Moving the Maxwell operator, $\mathcal{M}$, from the ket to bra part or vise-versa, by integration by parts produces a volume term proportional to the anti-Hermitian part of the material response, $\bar{\varepsilon}_I$, and a surface term carrying the Poynting flux. The resulting identity is the starting point for power balance and the optical theorem (Sec.~\ref{sec:energy_adjoint}).

In open systems, however, radiating fields decay only as $|\mathbf r|^{-1}$ at infinity and are therefore not square-integrable over all space. For compactly supported sources, the mixed inner product $\langle \mathcal S | \mathcal E\rangle$ is nevertheless well defined, since the integration is restricted to the finite support of $|\mathcal S\rangle$ and only requires $|\mathcal E\rangle$ to be locally square-integrable. 
We now develop the adjoint structure (Eq. \eqref{eq:adjoint}) of the Maxwell operator with respect to the $L^2$ inner product Eq.~\eqref{eq:L2ip}. For the material tensor $\bar{\varepsilon}$, which acts by pointwise multiplication, no integration by parts is needed:
\begin{equation}
\begin{split}
  \langle\mathcal{E}_1|
  \bar{\varepsilon}\,\mathcal{E}_2\rangle
  &= \int\mathcal{E}_1^\dagger\,\bar{\varepsilon}\,
    \mathcal{E}_2\,dV\\
  &= \int(\bar{\varepsilon}^\dagger\mathcal{E}_1)^\dagger
    \mathcal{E}_2\,dV
  = \langle\bar{\varepsilon}^\dagger\mathcal{E}_1|
    \mathcal{E}_2\rangle,
\end{split}
  \label{eq:eps_adjoint}
\end{equation}
so the formal adjoint is simply the pointwise Hermitian conjugate, with no boundary term. To separate the material response into Hermitian and anti-Hermitian components, we write
\begin{gather}
    \bar{\varepsilon} = \bar{\varepsilon}_R + i\bar{\varepsilon}_I,\notag\\
    \bar{\varepsilon}_R = \tfrac{1}{2}(\bar{\varepsilon}+\bar{\varepsilon}^\dagger),\quad
    \bar{\varepsilon}_I = \tfrac{1}{2i}(\bar{\varepsilon}-\bar{\varepsilon}^\dagger),
    \label{eq:eps_decomp}
\end{gather}
where $\bar{\varepsilon}_R$ and $\bar{\varepsilon}_I$ are both Hermitian.

% \paragraph{Differential operators.}

Integration by parts on the dual curl operator $\bar{\nabla}\times$ produces a boundary flux term that cannot be discarded in an open domain.
% The dual curl operator $\bar  
Using the vector Green's identity for dual fields we obtain (Appendix~\ref{app:vector_green}):
\begin{multline}
  \langle\bar{\nabla}\!\times\!\mathcal{E}_1|
    \mathcal{E}_2\rangle
  + \langle\mathcal{E}_1|
    \bar{\nabla}\!\times\!\mathcal{E}_2\rangle\\
  = -Z_0\oint\hat{\mathbf{n}}\!\cdot\!
    \bigl(\mathbf{E}_2\!\times\!\mathbf{H}_1^*
    + \mathbf{E}_1^*\!\times\!\mathbf{H}_2\bigr)\,dS.
  \label{eq:vector_green}
\end{multline}
%We note that this is the only use of vector calculus in this paper. 
To write this compactly, we introduce the dual surface operator. Given a surface $S$ enclosing the integration volume, with outward unit normal $\hat{\mathbf{n}}$, we define
\begin{equation}
  (\bar{n}\times)\,\mathcal{E}(\mathbf r)
  \equiv \begin{bmatrix}
    \hat{\mathbf{n}}\times(Z_0\mathbf{H}(\mathbf r))\\[2pt]
    -\hat{\mathbf{n}}\times\mathbf{E}(\mathbf r)
  \end{bmatrix},
  \label{eq:nbar_def}
\end{equation}
which extracts the tangential field components that appear in surface flux integrals. The associated surface pairing is
\begin{equation}
  \langle\mathcal{E}_1|
  \bar{n}\times\mathcal{E}_2\rangle_S
  \equiv \oint\mathcal{E}_1^\dagger\cdot
  (\bar{n}\times\mathcal{E}_2)\,dS.
  \label{eq:surface_ip}
\end{equation}
For $\mathcal E_1=\mathcal E_2=\mathcal E$, the surface term reduces to
$-2Z_0\,\mathrm{Re}\,[\hat{\mathbf n}\cdot(\mathbf E\times \mathbf H^*)]$,
which is $Z_0$ times the time-averaged Poynting flux through $S$.
% Expanding in components,
% \begin{equation}
%   \mathcal{E}_1^\dagger\cdot(\bar{n}\times\mathcal{E}_2)
%   = \mathbf{E}_1^*\!\cdot(\hat{\mathbf{n}}\times Z_0\mathbf{H}_2)
%   - (Z_0\mathbf{H}_1)^*\!\cdot(\hat{\mathbf{n}}\times\mathbf{E}_2),
%   \label{eq:surface_expand}
% \end{equation}
% which, by the scalar triple product identity, equals 
% $-Z_0\,\hat{\mathbf{n}}\cdot(\mathbf{E}_1^*\times\mathbf{H}_2 + \mathbf{E}_2\times\mathbf{H}_1^*).
% $
% When $\mathcal{E}_1 = \mathcal{E}_2 = \mathcal{E}$, this reduces to $-2Z_0\,\mathrm{Re}[\hat{\mathbf{n}}\cdot(\mathbf{E}\times\mathbf{H}^*)]$, which is $Z_0$ times the time-averaged Poynting flux through $S$. 
With this notation, the vector Green's identity takes the compact form
\begin{equation}
  \langle\bar{\nabla}\times\mathcal{E}_1|
    \mathcal{E}_2\rangle
  + \langle\mathcal{E}_1|
    \bar{\nabla}\times\mathcal{E}_2\rangle
  = \langle\mathcal{E}_1|
    \bar{n}\times\mathcal{E}_2\rangle_S,
  \label{eq:vector_green_compact}
\end{equation}
showing that $\bar{\nabla}\times$ is skew-adjoint, i.e., transferring it across the inner product flips its sign, with the mismatch carried by the surface term. The Maxwell Hamiltonian $\mathcal{H}=i\bar{\nabla}\times$ includes a factor of $i$ that converts skew-adjointness to self-adjointness, giving
\begin{equation}
  \langle\mathcal{E}_1|
  \mathcal{H}\mathcal{E}_2\rangle
  = \langle\mathcal{H}\mathcal{E}_1|
    \mathcal{E}_2\rangle
  + i\langle\mathcal{E}_1|
    \bar{n}\times\mathcal{E}_2\rangle_S,
  \label{eq:H_adjoint}
\end{equation}
so $\mathcal{H}^\dagger = \mathcal{H}$, with boundary form $\mathcal{B} = i\langle\mathcal{E}_1|\bar{n}\times\mathcal{E}_2\rangle_S$.

%\paragraph{Adjoint of the Maxwell operator.}
Combining Eqs.~\eqref{eq:H_adjoint} and~\eqref{eq:eps_adjoint} for $\mathcal{M} = \mathcal{H} - k_0\bar{\varepsilon}$ results in
\begin{multline}
\langle\mathcal{E}_1|\mathcal{M}\mathcal{E}_2\rangle
- \langle\mathcal{M}^\dagger\mathcal{E}_1|\mathcal{E}_2\rangle
=i\langle\mathcal{E}_1|\bar{n}\!\times\!\mathcal{E}_2\rangle_S.
\label{eq:m_adjoint_identity}
\end{multline}
and using $\bar{\varepsilon}^\dagger = \bar{\varepsilon} - 2i\bar{\varepsilon}_I$ with $\bar{\varepsilon}_I^\dagger = \bar{\varepsilon}_I$ yields the energy adjoint identity:
\begin{multline}
\langle\mathcal{E}_1|\mathcal{M}\mathcal{E}_2\rangle
- \langle\mathcal{M}\mathcal{E}_1|\mathcal{E}_2\rangle\\
= -2ik_0\langle\mathcal{E}_1|\bar{\varepsilon}_I\mathcal{E}_2\rangle
+ i\langle\mathcal{E}_1|\bar{n}\!\times\!\mathcal{E}_2\rangle_S.
\label{eq:energy_adjoint_identity}
\end{multline}
The left-hand side measures the failure of $\mathcal{M}$ to be self-adjoint. The right-hand side decomposes this failure into two physically distinct contributions: a volume term proportional to the anti-Hermitian part of the material response, $\bar{\varepsilon}_I$, which accounts for absorption or gain in the medium; and a surface term carrying the electromagnetic flux through the boundary $S$. When both vanish, i.e.,\ for a lossless medium with boundary conditions that eliminate the surface flux, the Maxwell operator is formally self-adjoint.

%\paragraph{Power balance.}
As a first application of the energy adjoint identity, we derive the power balance for a source-driven field. Choosing $\mathcal{E}_1=\mathcal{E}_2=\mathcal{E}$ and substituting $\mathcal{M}|\mathcal{E}\rangle=|i\mathcal{J}\rangle$ into Eq.~\eqref{eq:energy_adjoint_identity}, one obtains, upon taking the real part, Poynting's theorem in operator form:
\begin{equation}
\mathrm{Re}\,\langle\mathcal{J}|\mathcal{E}\rangle
= -k_0\langle\mathcal{E}|\bar{\varepsilon}_I\mathcal{E}\rangle
+ \tfrac{1}{2}\langle\mathcal{E}|\bar{n}\times\mathcal{E}\rangle_S,
\label{eq:poynting_operator}
\end{equation}
% Both terms on the right-hand side are real: the first because $\bar{\varepsilon}_I$ is Hermitian, and the second by Eq.~\eqref{eq:vector_green}.
%  Setting $\mathcal{E}_1 = \mathcal{E}_2 = \mathcal{E}$ with $\mathcal{M}|\mathcal{E}\rangle = |i\mathcal{J}\rangle$ in Eq.~\eqref{eq:energy_adjoint_identity} and taking real parts gives Poynting's theorem in operator form:
% \begin{equation}
% \mathrm{Re}\langle\mathcal{J}|\mathcal{E}\rangle
% = -k_0\langle\mathcal{E}|\bar{\varepsilon}_I\mathcal{E}\rangle
% + \tfrac{1}{2}\langle\mathcal{E}|\bar{n}\times\mathcal{E}\rangle_S,
% \label{eq:poynting_operator}
% \end{equation}
% where both terms on the right-hand side are real: $\langle\mathcal{E}|\bar{\varepsilon}_I\mathcal{E}\rangle$ because $\bar{\varepsilon}_I$ is Hermitian, and $\langle\mathcal{E}|\bar{n}\times\mathcal{E}\rangle_S$ as shown in Eq.~\eqref{eq:surface_expand}.

The left-hand side is proportional to the time-averaged power delivered by the sources. The first term on the right is the power dissipated in the medium, which is negative(positive) for a passive(active) material with $\bar{\varepsilon}_I > 0$($\bar{\varepsilon}_I<0$). The surface term is the net outward Poynting flux, likewise negative when energy radiates away from the sources. Negating both sides recovers the standard power budget: source power is equal to the dissipated power plus radiated power. The energy adjoint identity thus encodes the complete Poynting theorem in a single line; the same identity, applied to Green's functions rather than fields, will yield the generalized optical theorem in Sec.~\ref{sec:optical_theorem}.

%%%%%%%%%%%%%%%%%%%%%%%%%%%%%%%%%%%
\subsection{Adjoint of the Green's Operator}
\label{sec:greens_adjoint}
%%%%%%%%%%%%%%%%%%%%%%%%%%%%%%%%%%%
To formulate the generalized optical theorem, we first introduce the retarded Green operator associated with the Maxwell operator. Under outgoing boundary conditions, the retarded Green's operator is
\begin{equation}
G^{(+)}(\omega) = \lim_{\eta \to 0^+}\bigl(\mathcal{H} - k_0\bar{\varepsilon} - i\eta\bigr)^{-1},
\label{eq:G_retarded_def}
\end{equation}
where $-i\eta$ shifts the poles off the real axis, selecting the retarded solution. From now on, $G$ without a superscript denotes the retarded Green's operator $G^{(+)}$.
Unlike the differential operator $\mathcal{M}$, the Green's operator acts by integration rather than differentiation. For an integral operator such as $G$, the $L^2$ adjoint is obtained by exchanging the order of integration: it simultaneously conjugate-transposes the matrix kernel and swaps the spatial arguments, with no boundary term generated. This is in contrast to the differential operator $\mathcal{M}$, whose adjoint requires integration by parts and produces a surface contribution (Eq.~\eqref{eq:energy_adjoint_identity}). The absence of a boundary term in the adjoint of $G$ is what later allows us to move $G$ freely between bra and ket in the derivation of the optical theorem. Special care is required at the coincident-point singularity $\mathbf{r}=\mathbf{r}'$, discussed below. Formally, the adjoint is defined by
\begin{equation}
\langle\mathcal{S}_1|G\mathcal{S}_2\rangle = \langle G^\dagger\mathcal{S}_1|\mathcal{S}_2\rangle.
\label{eq:G_no_boundary}
\end{equation}
Taking the energy adjoint of the resolvent Eq.~\eqref{eq:G_retarded_def}, with $\mathcal{H}^\dagger = \mathcal{H}$ from Sec.~\ref{sec:energy_adjoint}, conjugate-transposes the material tensor and conjugates the scalar $-i\eta\to +i\eta$:
\begin{equation}
G^\dagger(\omega) = \bigl(\mathcal{H} - k_0\bar{\varepsilon}^\dagger + i\eta\bigr)^{-1}.
\label{eq:G_adjoint}
\end{equation}
Thus two changes occur simultaneously. First, the material response is replaced by its Hermitian conjugate,
$\bar{\varepsilon}^\dagger=\bar{\varepsilon}_R-i\bar{\varepsilon}_I$:
the reactive part is unchanged, while the dissipative part changes sign, interchanging absorption and gain. Second, the radiation prescription switches from outgoing to incoming, since the $+i\eta$ shift selects the advanced solution. Therefore $G^\dagger$ is the advanced Green operator of the adjoint medium. In the lossless case, $\bar{\varepsilon}_I=0$, so the medium itself is unchanged and $G^\dagger$ reduces to the advanced Green operator of the same medium. 
% Two things change simultaneously. First, the material response is replaced by its Hermitian conjugate $\bar{\varepsilon}^\dagger = \bar{\varepsilon}_R - i\bar{\varepsilon}_I$: the reactive part is unchanged, while the dissipative part reverses sign, interchanging absorption and gain. This is the material response of the time-reversed medium. Second, the boundary condition switches from outgoing to incoming: the $+i\eta$ prescription selects the advanced solution, so $G^\dagger$ propagates from future to source time rather than from excitation to future. In the lossless case ($\bar{\varepsilon}_I = 0$), the medium is its own time-reverse and $G^\dagger$ reduces to the advanced Green's operator for the same medium.

%\paragraph{Kernel-level identity.}
Next we look at the Green's operator in position representation. 
The Green's operator acts through a $6\times 6$ kernel,
\begin{equation}
\langle \mathbf r|G|\mathcal S\rangle = \int g(\mathbf{r},\mathbf{r}')\,\mathcal{S}(\mathbf{r}')\,dV'.
\label{eq:kernel_action}
\end{equation}
To obtain the adjoint kernel, we substitute this into the left-hand side of Eq.~\eqref{eq:G_no_boundary}:
\begin{equation}
\langle\mathcal{S}_1|G\mathcal{S}_2\rangle
= \int \mathcal{S}_1^\dagger(\mathbf{r})
\biggl(\int g(\mathbf{r},\mathbf{r}')\,
\mathcal{S}_2(\mathbf{r}')\,dV'\biggr)dV.
\label{eq:kernel_expand}
\end{equation}
Exchanging the order of integration and regrouping,
\begin{equation}
\begin{split}
&= \iint \bigl[g^{*T}(\mathbf{r},\mathbf{r}')\,
\mathcal{S}_1(\mathbf{r})\bigr]^\dagger
\mathcal{S}_2(\mathbf{r}')\,dV\,dV'\\
&= \int \biggl[\int g^{*T}(\mathbf{r},\mathbf{r}')\,
\mathcal{S}_1(\mathbf{r})\,dV\biggr]^\dagger
\mathcal{S}_2(\mathbf{r}')\,dV',
\end{split}
\label{eq:kernel_regroup}
\end{equation}
where $*T$ denotes the conjugate transpose of the $6\times 6$ matrix. Comparing with $\langle G^\dagger\mathcal{S}_1|\mathcal{S}_2\rangle = \int(G^\dagger\mathcal{S}_1)^\dagger(\mathbf{r}')\,\mathcal{S}_2(\mathbf{r}')\,dV'$, we identify the adjoint kernel action as
\begin{equation}
\langle \mathbf r'|G^\dagger|\mathcal S_1\rangle = \int g^\dagger(\mathbf{r}',\mathbf{r})\,\mathcal{S}_1(\mathbf{r})\,dV,
\label{eq:G_adjoint_kernel_action}
\end{equation}
with the pointwise matrix adjoint
\begin{equation}
\bigl(g(\mathbf{r},\mathbf{r}')\bigr)^\dagger = g^{*T}(\mathbf{r},\mathbf{r}')
\label{eq:kernel_adjoint}
\end{equation}

To summarize: the $L^2$ adjoint of the Green operator simultaneously conjugate-transposes the $6\times 6$ matrix and exchanges the spatial arguments. The kernel of $G^\dagger$ evaluated at $(\mathbf{r}',\mathbf{r})$ is the pointwise conjugate-transpose of $g$ at $(\mathbf{r},\mathbf{r}')$: in index notation, $[G^\dagger]_{ij}(\mathbf{r}',\mathbf{r}) = g_{ji}^*(\mathbf{r},\mathbf{r}')$.

\paragraph{Green kernel singularity.}
The exchange of integration order in Eq.~\eqref{eq:kernel_expand} requires justification, since the Green's kernel has a nonintegrable $|\mathbf{r}-\mathbf{r}'|^{-3}$ singularity near the source region. The standard decomposition of $g = \mathrm{PV}\,g + g_{\mathrm{ct}}$ into a principal-value part and a contact (delta-function) part handles this. The contact term yields, in the limit of an infinitesimal exclusion volume about the source point, a volume-independent depolarization dyad~\cite{Yaghjian1980}. The remaining contribution is understood in the principal-value sense, which makes the integral well defined for smooth, compactly supported sources evaluated outside the exclusion volume and thereby justifies the interchange of integrations. Since both contributions separately satisfy Eq.~\eqref{eq:kernel_adjoint}, the identity follows for the full kernel.
\subsection{Interior Field Representation}
\label{sec:stratton_chu_l2}
We apply the adjoint identity of $\mathcal{M}$, Eq.~\eqref{eq:m_adjoint_identity},
to derive a representation of
the field at any interior point of a finite region in terms of volume sources and the
tangential boundary data (Eq.~\eqref{eq:stratton_chu_intro}).
In the quantum theory this surface term acquires an independent
role as a channel for incoming vacuum fluctuations, which motivates
keeping it explicit throughout.

To obtain a pointwise field representation, we test the adjoint identity
with a source localized at the observation point $\mathbf r$. 
% To extract the field value at a specific interior point $\mathbf{r}$
% from the adjoint identity, we need a test source whose inner product
% with $\mathcal{E}_2$ returns $\mathcal{E}_2(\mathbf{r})$. 
% This
% requires a test field whose action under $\mathcal{M}^\dagger$ is 
% a
% delta function at $\mathbf{r}$, so that
% $\langle\mathcal{M}^\dagger\mathcal{E}_1|\mathcal{E}_2\rangle =
% \mathcal{E}_2(\mathbf{r})$. The field with this property is
% $G^\dagger\delta^{(3)}(\mathbf{r}-\cdot)$, so we set
% $|\mathcal{S}_1\rangle = |\delta^{(3)}(\mathbf{r}-\cdot)I_6\rangle$ and choose
% \jon{$|\delta^{(3)} \rangle$}  
For this, we choose
\begin{equation}
|\mathcal{E}_1\rangle = G^\dagger|\mathcal{S}_1\rangle,
\label{eq:test_field}
\end{equation}
which satisfies $\mathcal{M}^\dagger|\mathcal{E}_1\rangle =
|\mathcal{S}_1\rangle$ exactly since $\mathcal{M}^\dagger G^\dagger =
I$. Here $|\mathcal{S}_1\rangle$ is arbitrary; since the final result
must hold for all $|\mathcal{S}_1\rangle$, we can factor it out at the
end to obtain a pointwise identity for $\mathcal{E}_2$ directly. Substituting~\eqref{eq:test_field} into the adjoint
identity~\eqref{eq:energy_adjoint_identity} and using
$\mathcal{M}|\mathcal{E}_2\rangle = |\mathcal{S}_2\rangle$ we get:
\begin{equation}
\langle G^\dagger\mathcal{S}_1|\mathcal{S}_2\rangle
- \langle\mathcal{S}_1|\mathcal{E}_2\rangle
= i\langle G^\dagger\mathcal{S}_1|\bar{n}\times\mathcal{E}_2\rangle_S.
\label{eq:sc_intermediate}
\end{equation}
Since the integral operator $G$ generates no boundary term under the
$L^2$ inner product (Eq.~\eqref{eq:G_no_boundary}), $G^\dagger$ can
be moved freely from bra to ket:
\begin{align}
\langle G^\dagger\mathcal{S}_1|\mathcal{S}_2\rangle
&= \langle\mathcal{S}_1|G\mathcal{S}_2\rangle,
\nonumber\\
\langle G^\dagger\mathcal{S}_1|\bar{n}\times\mathcal{E}_2\rangle_S
&= \langle\mathcal{S}_1|G(\bar{n}\times\mathcal{E}_2)\rangle_S.
\label{eq:G_move}
\end{align}
Once $G^\dagger$ is moved to the ket, the operator acting is the
original retarded resolvent $G = (\mathcal{H} - k_0\bar{\varepsilon}
- i\eta)^{-1}$ in the original medium with outgoing boundary
conditions. The adjoint medium $\bar{\varepsilon}^\dagger$ and the incoming character of $G^\dagger$ appear only at an intermediate stage of the derivation. This is consistent since the adjoint identity is a purely algebraic consequence of integration by parts and does not require the two fields to satisfy the same boundary conditions or belong to the same operator domain. This is analogous to the elementary identity
\begin{equation}
\int u\,(\partial_x v)\,dx
=
-\int (\partial_x u)\,v\,dx
+
\bigl[u\,v\bigr]_{\partial},
\end{equation}
which holds for arbitrary sufficiently regular $u$ and $v$, regardless of whether they obey the same boundary conditions. 
Substituting~\eqref{eq:G_move} into~\eqref{eq:sc_intermediate} and 
factoring out $\langle\mathcal{S}_1|$
% , and evaluating at
% $\mathcal{S}_1 = |\delta^{(3)}\rangle$ 
gives the
first-order field representation in position space
\begin{multline}
    \mathcal{E}(\mathbf{r}) = \int_V g(\mathbf{r},\mathbf{r}')\,
\mathcal{S}(\mathbf{r}')\,dV'
\\- i\oint_S g(\mathbf{r},\mathbf{s})\,(\bar{n}\times)\,
\mathcal{E}(\mathbf{s})\,dS.
\label{eq:stratton_chu_first_order}
\end{multline}

% \paragraph{Remark on generality.}
This result holds for any inner product under which
$\mathcal H=i\bar\nabla\times$ is formally self-adjoint, i.e.\ any
pairing whose integration by parts produces only a surface term with no
bulk residual. The surface term in
Eq.~\eqref{eq:energy_adjoint_identity} is generated entirely by
$\mathcal H$; the material part $k_0\bar\varepsilon$ acts by pointwise
multiplication and contributes nothing to the concomitant thereby generalising Eq.~\eqref{eq:stratton_chu_first_order} to arbitrary linear media.
% % The surface
% term in Eq.~\eqref{eq:energy_adjoint_identity} is generated entirely
% by $\mathcal{H}$; the material part $k_0\bar{\varepsilon}$
% acts by pointwise multiplication and contributes nothing to
% the concomitant.  
% Moving $G^\dagger$ from bra to ket in Eq.~\eqref{eq:G_move}
% uses the kernel identity~\eqref{eq:kernel_adjoint}: \[
% g^\dagger(\mathbf r',\mathbf r)=g(\mathbf r,\mathbf r')^\dagger,
% \]
% which is the position-space statement of the adjoint action. The operator acting on the
% ket is therefore once again the original retarded resolvent
% $
% G=(\mathcal H-k_0\bar\varepsilon-i\eta)^{-1},
% $
% with the original medium $\bar\varepsilon$ and outgoing boundary
% condition.  Thus the adjoint medium $\bar\varepsilon^\dagger$ and the
% incoming character of $G^\dagger$ appear only at the intermediate
% stage of the derivation and do not survive in the final formula.

%%%%%%%%%%%%%%%%%%%%%%%%%%%%%%%%%%%
\subsection{Generalized Optical Theorem}
\label{sec:optical_theorem}
%%%%%%%%%%%%%%%%%%%%%%%%%%%%%%%%%%%
\begin{figure}
    \centering
    \includegraphics[width=1.\linewidth]{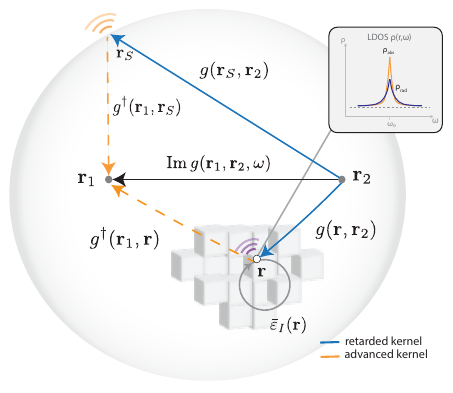}
    \caption{\textbf{Surface-volume decomposition of the Green tensor}. The direct response between $\mathbf r_1$ and $\mathbf r_2$ is related to two mediated pathways: a volume contribution through an interior point $\mathbf r$ in the lossy region $\bar{\varepsilon}_I(\mathbf r)$, and a surface contribution through a boundary point $\mathbf r_S$. The inset indicates the LDOS at $\mathbf r$.}
    \label{fig:optical_theorem}
\end{figure}
% \begin{figure}
%     \centering
%     \includegraphics[width=1.\linewidth]{figures/fig2_ot2.pdf}
%     \caption{Diagrammatic form of the optical-theorem identity. The difference between retarded and advanced $g(\mathbf r_1,\mathbf r_2)$ is written as the sum of a volume term, mediated by an interior lossy point weighted by $\bar{\varepsilon}_I(\mathbf r)$, and a surface term, mediated by a boundary point $\mathbf r_S$.}
%     \label{fig:placeholder}
% \end{figure}
We now specialize the energy adjoint identity to retarded fields, obtaining a constraint on the Green's operator itself that generalizes the optical theorem of scattering theory~\cite{newton1976optical,newton1982scattering}. Substituting $|\mathcal{E}_1\rangle = G|\mathcal{S}_1\rangle$ and $|\mathcal{E}_2\rangle = G|\mathcal{S}_2\rangle$ into Eq.~\eqref{eq:energy_adjoint_identity}, with $\mathcal{M}|\mathcal{E}_{1,2}\rangle = |\mathcal{S}_{1,2}\rangle$, each term acquires $G$ inside the bra. Since the $L^2$ adjoint of $G$ generates no boundary term (Eq.~\eqref{eq:G_no_boundary}), every occurrence can be moved to the ket as $G^\dagger$, and because the result holds for arbitrary sources we can read off a kernel identity. The generalized optical theorem reads
\begin{equation}
\begin{split}
g(\mathbf{r}_1,\mathbf{r}_2) &- g^\dagger(\mathbf{r}_1,\mathbf{r}_2)\\
&= 2ik_0 \int g^\dagger(\mathbf{r}_1,\mathbf{r})\bar{\varepsilon}_I(\mathbf{r})\, g(\mathbf{r},\mathbf{r}_2) \, dV \\
&\quad -  i\oint g^\dagger(\mathbf{r}_1,\mathbf{r}_S)\,(\bar{n}\times)\, g(\mathbf{r}_S,\mathbf{r}_2) \, dS.
\end{split}
\label{eq:optical_theorem_kernel}
\end{equation}

Using the kernel adjoint relation~\eqref{eq:kernel_adjoint}, the left-hand side is the anti-Hermitian part of the Green's kernel: 
\begin{equation}
     g(\mathbf r_1, \mathbf r_2) - g^\dagger(\mathbf r_1, \mathbf r_2)= 2i\, \mathrm{Im}\,g (\mathbf r_1, \mathbf r_2).
\end{equation} 

The right-hand side decomposes it into the same two dissipation channels that appeared in the power balance (Fig.~\ref{fig:optical_theorem}): a volume integral weighted by the absorptive part of the material response $\bar{\varepsilon}_I$, and a surface integral encoding the radiated flux through $S$. 
% The identity now constrains the propagator rather than the fields.
Written as an operator identity, Eq.~\eqref{eq:optical_theorem_kernel} takes the compact form
\begin{multline}
G - G^\dagger = 2ik_0\,G^\dagger\,\bar{\varepsilon}_I\,G
- i\,G^\dagger\,(\bar{n}\times)\,G,
\label{eq:optical_theorem_operator}
\end{multline}
where $(\bar{n}\times)$ denotes the boundary operator acting on fields
restricted to $S$ and all operators act to the right within the inner product $\langle\mathcal{S}_1|(\cdots)\mathcal{S}_2\rangle$.

\paragraph{Connection to the algebraic resolvent identity.}
For bounded operators on a Hilbert space, the resolvent identity $A^{-1} - B^{-1} = A^{-1}(B-A)B^{-1}$ holds exactly. Our operators are unbounded and generate surface terms under the inner product, so the identity captures only the bulk contribution. Setting $A = \mathcal{M}$ and $B = \mathcal{M}^\dagger = \mathcal{M} + 2ik_0\bar{\varepsilon}_I$ gives
\begin{equation}
G - G^\dagger = 2ik_0\,G^\dagger\,\bar{\varepsilon}_I\,G,
\label{eq:resolvent_identity}
\end{equation}
while setting $A = \mathcal{M}^\dagger$ and $B = \mathcal{M}$ yields the alternative form
\begin{equation}
G - G^\dagger = 2ik_0\,G\,\bar{\varepsilon}_I\,G^\dagger.
\label{eq:resolvent_identity_alt}
\end{equation}
Both forms appear in subsequent derivations. In an open system, neither is complete: $G\mathcal{M} \neq \mathbb{I}$ for fields that do not satisfy the outgoing boundary condition, and the discrepancy is precisely the boundary-flux term of the full optical theorem Eq.~\eqref{eq:optical_theorem_operator}. In a fully absorptive medium ($\bar{\varepsilon}_I > 0$ everywhere) or closed system, the retarded fields decay at infinity, the surface term vanishes, and the resolvent identity and the optical theorem coincide. The energy-adjoint derivation, which keeps the boundary term explicit from the outset, remains valid in both regimes.

%%%%%%%%%%%%%%%%%%%%%%%%%%%%%%%%%%%%%%%%%%%%%%%%%%%
\subsection{Reciprocal pairing and adjoint identity}
\label{sec:reciprocity}

As a further application of using adjoint spaces to derive boundary identities, we introduce a second inner product under which the same adjoint procedure yields Lorentz reciprocity and the Green-function symmetry relation.

Lorentz reciprocity is the symmetry under which a source and detector can be exchanged with identical results. It breaks down in nonreciprocal systems such as magneto-optic materials. In the first-order formalism, swapping source and detector also reverses the propagation direction, which flips the sign of the magnetic field. We account for this by introducing the field-flip operator
\begin{equation}
\Pi = \mathrm{diag}(\mathbf{I}_3, -\mathbf{I}_3),
\end{equation}
which acts as $\Pi\mathcal{E} = [\mathbf{E}, -Z_0\mathbf{H}]^T$. The reciprocity identity for the Green kernel then takes the form
\begin{equation}
g(\mathbf{r}_1, \mathbf{r}_2) = \Pi \, g^T(\mathbf{r}_2, \mathbf{r}_1)\, \Pi.
\label{eq:green_reciprocity}
\end{equation}
This corresponds to the reciprocity symmetry of swapping the source and detector. 

To derive this, we define the reciprocal inner product
\begin{equation}
\langle\mathcal{E}_1|\mathcal{E}_2\rangle_R = \int_V \mathcal{E}_1^T \, \Pi \, \mathcal{E}_2 \, dV.
\label{eq:reciprocal_ip}
\end{equation}
Mathematically, this is a bilinear pairing under which $\mathcal{M}$ is formally self-adjoint (for reciprocal media) and $G$ is self-adjoint, with no boundary term for the integral operator.
For an inner product with just transpose, the $\mathcal{H}$ operator formal adjoint is $-\mathcal{H}$.  The $\Pi$-weight makes it self-adjoint, playing a similar role that the $i$ did in $\mathcal{H}$ for the energy adjoint. 
%The role of $\Pi$ is analogous to the factor of $i$ in the energy inner product: it flips the sign of the magnetic component. 

Using this inner product, we show in Appendix~\ref{app:reciprocal_adjoint} that the reciprocal adjoint of $G$,  defined as $G^\sharp$, satisfies $G^\sharp = G$ for reciprocal media ($\bar{\varepsilon}^T = \bar{\varepsilon}$). The kernel relation $g^\sharp(\mathbf{r},\mathbf{r}') = \Pi \, g^T(\mathbf{r}',\mathbf{r}) \, \Pi$, directly yielding Eq.~\eqref{eq:green_reciprocity}. The reciprocal adjoint identity
\begin{equation}
\langle\mathcal{E}_1|\mathcal{M}\mathcal{E}_2\rangle_R - \langle\mathcal{M}\mathcal{E}_1|\mathcal{E}2\rangle_R = i\langle\mathcal{E}1|\bar{n}\times\mathcal{E}2\rangle_{R,S}
\label{eq:reciprocal_adjoint_identity}
\end{equation}
has no volume term — material loss does not spoil reciprocity. Substituting $\mathcal{M}\mathcal{E}_{1,2} = \mathcal{S}_{1,2}$ into this identity yields Lorentz reciprocity:
\begin{equation}
\langle\mathcal{S}_1|\mathcal{E}_2\rangle_R - \langle\mathcal{E}_1|\mathcal{S}_2\rangle_R = i\langle\mathcal{E}_1|\bar{n}\times\mathcal{E}2\rangle_{R,S}.
\label{eq:lorentz_reciprocity_first_order}
\end{equation}

Substituting $\mathcal{M}\mathcal{E}_{1,2}=\mathcal{S}_{1,2}$ into Eq.~\eqref{eq:reciprocal_adjoint_identity} and expanding the dual-field products into their $3$-vector components yields the impedance-normalized Lorentz reciprocity relation,
\begin{equation}
\begin{split}
\int\Big[
  &Z_0\mathbf{J}_{E,1}\cdot\mathbf{E}_2
  - Z_0\mathbf{J}_{E,2}\cdot\mathbf{E}_1\\
  -\;&\mathbf{J}_{M,1}\cdot Z_0\mathbf{H}_2
  + \mathbf{J}_{M,2}\cdot Z_0\mathbf{H}_1
\Big]\,dV\\
&= iZ_0\oint\hat{\mathbf{n}}\cdot\bigl(
     \mathbf{E}_1\times\mathbf{H}_2
   - \mathbf{E}_2\times\mathbf{H}_1\bigr)\,dS.
\end{split}
\label{eq:lorentz_reciprocity}
\end{equation}

Physically, the reciprocal pairing can also be interpreted as the overlap of a field with its time-reversed counterpart, with the time reversal operator 
\begin{equation}
\mathcal{T}\mathcal{E}(\mathbf{r},\omega)
= \Pi\,\mathcal{E}^*(\mathbf{r},\omega)
= \begin{bmatrix}
    \mathbf{E}^*(\mathbf{r},\omega)\\[2pt] -Z_0\mathbf{H}^*(\mathbf{r},\omega)
  \end{bmatrix}.
\label{eq:time_reversal_def}
\end{equation}

\section{Plane-to-plane propagation and transfer operators}
\label{sec:transfer}

The first-order propagation formula of Sec.~\ref{sec:stratton_chu_l2} directly yields a key transfer-matrix relation.
% With the first-order framework established, we glance over one of the crucial transfer matrix relation stemming from the first-order Stratton--Chu formula discussed in Sec.~\ref{sec:stratton_chu_l2}. 
Eq. \eqref{eq:stratton_chu_intro} can be read not only as a representation of an existing solution, but also as an exact propagation rule between surfaces considered inside the arbitrary volume depicted in Fig.~\ref{fig:field_geometry}.

\begin{figure}
    \centering
    \includegraphics[width=1.0\linewidth]{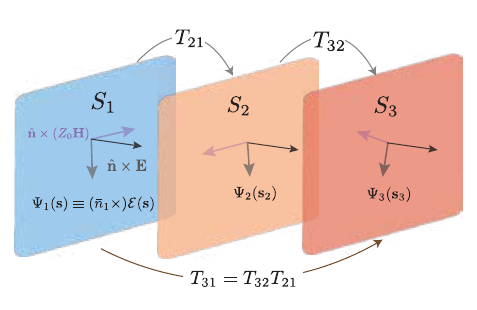}
    \caption{\textbf{Plane-to-plane transfer of the tangential electromagnetic state}. The trace $|\Psi_1\rangle$ on $S_1$ propagates to $|\Psi_2\rangle$ on $S_2$ through the transfer operator $T_{21}$, and then to $|\Psi_3\rangle$ on $S_3$ through $T_{32}$. The net propagation from $S_1$ to $S_3$ is therefore given by the exact composition law $T_{31}=T_{32}T_{21}$.}
    \label{fig:transfer}
\end{figure}
We consider three planes $S_1$, $S_2$, and $S_3$
 (or, more generally, three nonintersecting transverse surfaces) ordered along the propagation direction, as shown in Fig.~\ref{fig:transfer}. The regions between successive surfaces are assumed to be source-free.
% To understand the intermediate propagation process, let us consider $S_1$, $S_2$, and $S_3$ be three planes (or more generally, three nonintersecting transverse surfaces) ordered along the propagation direction, and suppose that the regions between successive surfaces are source-free, refer to Fig.~\ref{fig:transfer}. 
The tangential dual trace on each surface is written as 
\begin{equation}
\Psi_i(\mathbf{s}) \equiv (\bar{n}_i\times)\mathcal E(\mathbf{s}),
\qquad \mathbf{s}\in S_i.
\label{eq:tangential_trace_def}
\end{equation}

The propagation equation then relates the field at a new surface with $\mathbf{r}$ by
\begin{equation}
\mathcal E(\mathbf r)
=
-i\oint_{S_i} g^{(+)}(\mathbf r,\mathbf s)\,\Psi_i(\mathbf s)\,dS.
\label{eq:plane_propagation}
\end{equation}
 Taking the tangential trace on a second surface $S_j$ defines the exact surface-to-surface transfer operator
\begin{equation}
\Psi_j(\mathbf s_j)
=\!
\oint_{S_i}\!
T_{ji}(\mathbf s_j,\mathbf s_i)\,
\Psi_i(\mathbf s_i)\,dS_i,
\label{eq:Tji_def}
\end{equation}
with kernel
\begin{equation}
T_{ji}(\mathbf s_j,\mathbf s_i)
\equiv
-i(\bar{n}_j\times)\,
g^{(+)}(\mathbf s_j,\mathbf s_i).
\label{eq:Tji_kernel}
\end{equation}
This is the first-order analog of a transfer matrix: it propagates the full tangential electromagnetic state from one plane to the next.

The transfer operators compose exactly.  If $|\Psi_2\rangle=T_{21}|\Psi_1\rangle$ and $|\Psi_3\rangle=T_{32}|\Psi_2\rangle$, then
\begin{equation}
|\Psi_3\rangle = T_{32}T_{21}|\Psi_1\rangle,
\end{equation}
so
\begin{equation}
T_{31}=T_{32}T_{21}.
\label{eq:T_compose}
\end{equation}
In kernel form this reads
\begin{equation}
T_{31}(\mathbf s_3,\mathbf s_1)
=
\oint_{S_2}
T_{32}(\mathbf s_3,\mathbf s_2)\,
T_{21}(\mathbf s_2,\mathbf s_1)\,dS_2,
\label{eq:T_compose_kernel}
\end{equation}
which, when expressed back in terms of the Green's kernel, the same identity becomes the first-order Huygens composition rule
\begin{multline}
g^{(+)}(\mathbf r_3,\mathbf r_1)
=
\oint_{S_2}\!
g^{(+)}(\mathbf r_3,\mathbf s_2)\\
\cdot(\bar{n}_2\!\times)\,
g^{(+)}(\mathbf s_2,\mathbf r_1)\,dS_2,
\label{eq:g_compose}
\end{multline}
valid when $\mathbf{r}_3$ and $\mathbf{r}_1$ lie on opposite sides of $S_2$ and the region between them is source free with the same retarded boundary condition used throughout.
The exact composition law is therefore not a pointwise product of Green's kernels, but a surface-mediated convolution with the tangential trace operator $(\bar{n}\times)$ acting at the intermediate surface.

In a planar geometry with translational invariance, the surface integrals reduce to convolutions in the transverse coordinates, and after the Fourier transform, $T_{ji}$ becomes a $6\times 6$ matrix at each transverse wavevector, recovering the standard transfer-matrix formalism as a special case.

%%%%%%%%%%%%%%%%%%%%%%%%%%%%%%%%%%%%%%%%%%%%%
\section{Quantization in Dispersive-Absorptive Media}
\label{sec:quantization}
%%%%%%%%%%%%%%%%%%%%%%%%%%%%%%%%%%%%%%%%%%%%%
\subsection{Introduction}
In the preceding sections we derived a propagation equation expressing the dual field inside a region $V$ in terms of volumetric sources and the incoming boundary field.
% ~\eqref{eq:stratton_chu_intro}:
% \begin{equation}
% \begin{split}
% \mathcal{E}(\mathbf{r})
% ={}&\int g(\mathbf{r},\mathbf{r}')\,\mathcal{S}(\mathbf{r}')\,dV'\\
% &-i\oint g(\mathbf{r},\mathbf s)\,(\bar{n}\!\times)\,\mathcal{E}_{\mathrm{in}}(\mathbf s)\,dS.
% \end{split}
% \label{eq:classical_decomp}
% \end{equation}

We now work to promote this classical identity, Eq. \eqref{eq:stratton_chu_intro} to a quantum operator equation.
In a dispersive, absorptive medium, direct quantization of the electromagnetic field fails: the non-Hermiticity introduced by material loss and radiation causes mode amplitudes to decay, and the equal-time commutation relations are not preserved.

The resolution is a macroscopic quantization scheme in which the medium polarization and an absorptive reservoir are treated as dynamical quantum degrees of freedom alongside the radiation field, with the boundary carrying incoming vacuum fluctuations.
The resulting operator equation retains the structure of Eq.~\eqref{eq:stratton_chu_intro},  but the source is replaced by a Langevin noise operator whose strength is dictated by the local dissipation, ensuring that the commutation relations are maintained despite the loss.
We show that the quantized field exactly satisfies the fluctuation-dissipation theorem. The equal-frequency commutator closes through the generalized optical theorem of Sec.~\ref{sec:optical_theorem}, confirming that the quantization is self-consistent and unitary.

\subsection{Background}
The quantization of the electromagnetic field in dielectric media has a long history.
In lossless media, it can be developed by treating the material as a collection of polarizable oscillators coupled to the radiation field~\cite{glauber1991quantum, knoll1987action}.
Huttner and Barnett~\cite{huttner1992quantization} extended this picture to absorptive dielectrics by coupling each polarization oscillator to a continuum of bosonic bath modes representing microscopic loss channels, such as phonon emission and electronic relaxation.
Diagonalizing the resulting field--matter--bath Hamiltonian yields polaritonic ladder operators $\hat{\mathbf{f}}(\mathbf{r},\omega)$ that mix photonic, material, and reservoir character.

Gruner and Welsch~\cite{gruner1996green-function} reformulated this theory directly in terms of the classical dyadic Green's function, expressing the quantized electric field as a volume integral of bosonic noise operators weighted by $\mathrm{Im}\,\varepsilon$ and propagated by the Green tensor, which is now the standard approach in macroscopic QED~\cite{buhmann2007dispersion,scheel2006quantum}.

However, Drezet~\cite{drezet2017quantizing, drezet2017equivalence}, Ciattoni~\cite{ciattoni20240719quantum, ciattoni2025quantum-optical}, and Stefano~\cite{stefano2001mode} pointed out that for spatially finite objects the volume contribution alone is incomplete: vacuum fluctuations entering through the boundary, the ``boundary-assisted'' contribution, must also be retained to preserve the canonical commutation relations.

An equivalent route to the same physics is provided by the Heisenberg--Langevin approach~\cite{rosa2010quantum}: rather than diagonalizing the full Hamiltonian, one formally integrates out the reservoir and obtains a quantum Langevin equation for the polarization, with a noise operator fixed by the fluctuation-dissipation theorem.
We adopt this strategy here because it interfaces naturally with the first-order Green-operator formalism developed above, and because it keeps both fluctuation channels explicit: bulk material noise and incoming boundary vacuum fluctuations.

\subsection{Heisenberg-Langevin Equations}
\begin{figure}
    \centering
\includegraphics[width=1.0\linewidth]{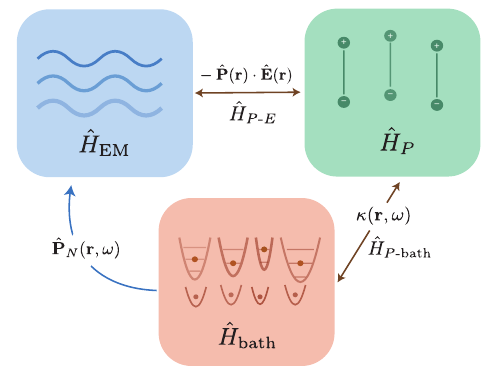}
    \caption{\textbf{Quantization scheme}. The total Hamiltonian couples three
subsystems: the free electromagnetic field $\hat{H}_{\mathrm{EM}}$, the
material polarization $\hat{H}_P$, and the bosonic bath $\hat{H}_{\mathrm{bath}}$.
The field and polarization interact via the dipole coupling
$-\hat{\mathbf{P}}(\mathbf{r})\cdot\hat{\mathbf{E}}(\mathbf{r})$, while the
polarization and bath are coupled with spectrum $\kappa(\mathbf{r},\omega)$.
The field and bath are never coupled directly.}
    \label{fig:quantization}
\end{figure}

\paragraph{Hamiltonian and constitutive relation.} We model the dispersive, absorptive medium by coupling the electromagnetic field to a
material polarization, which in turn is coupled to a local bosonic reservoir responsible
for dissipation. The total Hamiltonian is
\begin{equation}
\hat{H}
= \hat{H}_{\mathrm{EM}}
+ \hat{H}_{\mathcal{P}}
+ \hat{H}_{\mathrm{bath}}
+ \hat{H}_{\mathcal{P}\text{-}\mathcal{E}}
+ \hat{H}_{\mathcal{P}\text{-bath}},
\label{eq:H_total}
\end{equation}
where $\hat{H}_{\mathrm{EM}} = \frac{\varepsilon_0}{2}\int
  \bigl|\hat{\mathcal{E}}(\mathbf{r})\bigr|^2 d^3r$
is the free-field energy and $\hat{H}_{\mathcal{P}}$ describes harmonic oscillators
for both the electric polarization and the magnetization, with spatially varying
resonance frequencies $\omega_{0e}(\mathbf{r})$ and $\omega_{0m}(\mathbf{r})$.
The two sectors are independent and treated on equal footing, so all expressions
below are written for the dual polarization vector
\begin{equation}
\hat{\mathcal{P}} \equiv
\begin{bmatrix}
\hat{\mathbf{P}}/\varepsilon_0 \\ Z_0\hat{\mathbf{M}}
\end{bmatrix},
\label{eq:dual_polarization_def}
\end{equation}
whose components carry the same dimensions as $\hat{\mathcal{E}}$, and for the
six-component dual bath $\hat{\mathbf{b}} = [\hat{\mathbf{b}}_e, \hat{\mathbf{b}}_m]^T$
combining both reservoirs.
The dual bath Hamiltonian is
\begin{equation}
\hat{H}_{\mathrm{bath}}
= \int dV\int_0^\infty\!d\omega\;\hbar\omega\,
\hat{\mathbf{b}}^\dagger(\mathbf{r},\omega)\cdot
\hat{\mathbf{b}}(\mathbf{r},\omega),
\label{eq:H_bath}
\end{equation}
satisfying $[\hat{b}_\alpha(\mathbf{r},\omega),\,\hat{b}_\beta^\dagger(\mathbf{r}',\omega')]
= \delta_{\alpha\beta}\,\delta(\mathbf{r}-\mathbf{r}')\,\delta(\omega-\omega')$,
with cross-sector commutators $[\hat{b}_{e,i}, \hat{b}_{m,j}^\dagger] = 0$
reflecting the independence of the two reservoirs.
Physically, the bath modes at $\mathbf{r}$ represent all microscopic channels through
which electromagnetic energy stored at that point is irreversibly lost: phonon emission,
electronic relaxation, radiative cascades into dark modes, and so on.
The field--matter interaction is the dipole coupling
\begin{equation}
\hat{H}_{\mathcal{P}\text{-}\mathcal{E}}
= -\varepsilon_0\int\hat{\mathcal{P}}^\dagger\cdot\hat{\mathcal{E}}\,dV
\end{equation}
in the multipolar (Power--Zienau--Woolley~\cite{power_zienau,woolley2020power}) form,
and $\hat{H}_{\mathcal{P}\text{-bath}}$ couples the dual polarization linearly to
$\hat{\mathbf{b}}$ and $\hat{\mathbf{b}}^\dagger$ with block-diagonal coupling spectrum
$\bar{\kappa}(\mathbf{r},\omega)$.
The dual field and the dual bath are never coupled directly and all dissipation is
mediated by $\hat{\mathcal{P}}$(see Fig.~\ref{fig:quantization}).

Deriving the Heisenberg equations of motion, formally integrating out both
reservoirs simultaneously, and taking the Markov limit (Appendix~\ref{app:quantization})
yields the frequency-domain constitutive relation
\begin{equation}
\hat{\mathcal{P}}(\mathbf{r},\omega)
= \bar{\chi}(\mathbf{r},\omega)\,\hat{\mathcal{E}}(\mathbf{r},\omega)
+ \hat{\mathcal{P}}_N(\mathbf{r},\omega),
\label{eq:constitutive_quantum}
\end{equation}
where $\bar{\chi}$ is the dimensionless $6\times 6$ susceptibility tensor encoding
both electric and magnetic response and
$\hat{\mathcal{P}}_N = (\hat{\mathbf{P}}_N/\varepsilon_0,\;Z_0\hat{\mathbf{M}}_N)^T$
is the impedance-normalized Langevin noise polarization driven by the dual bath operators.
The fluctuation-dissipation theorem fixes the noise commutator to be proportional
to the local dissipation:
\begin{multline}
\bigl[\hat{\mathcal{P}}_{N,i}(\mathbf{r},\omega),\,
\hat{\mathcal{P}}_{N,j}^\dagger(\mathbf{r}',\omega')\bigr]\\
= \frac{\hbar}{\pi\varepsilon_0}\,(\bar{\varepsilon}_I)_{ij}(\mathbf{r},\omega)\,
\delta(\mathbf{r}-\mathbf{r}')\,\delta(\omega-\omega'),
\label{eq:FDT_dualP}
\end{multline}
where $\bar{\varepsilon}_I$ is the anti-Hermitian part of the dimensionless dual
material tensor from Eq.~\eqref{eq:eps_decomp}, with cross-sector commutators
vanishing identically because the electric and magnetic reservoirs are independent.
Each spatial point radiates noise in exact proportion to the energy it absorbs,
ensuring thermodynamic consistency~\cite{franke2020fluctuation}.

% \begin{figure}
%     \centering
%     \includegraphics[width=1.0\linewidth]{figures/fig2_quantum.pdf}
%     \caption{ \jon{make a little dark, change dipoles to lines} Quantization scheme. (a) The total Hamiltonian couples three
% subsystems: the free electromagnetic field $\hat{H}_{\mathrm{EM}}$, the
% material polarization $\hat{H}_P$, and the bosonic bath $\hat{H}_{\mathrm{bath}}$.
% The field and polarization interact via the dipole coupling
% $-\hat{\mathbf{P}}(\mathbf{r})\cdot\hat{\mathbf{E}}(\mathbf{r})$, while the
% polarization and bath are coupled with spectrum $\kappa(\mathbf{r},\omega)$.
% The field and bath are never coupled directly. (b) After integrating out the
% bath in the Markov limit, the polarization acquires a Langevin noise operator
% $\hat{\mathcal{P}}_N(\mathbf{r}')$ whose commutator is fixed by the
% fluctuation-dissipation theorem. The total quantized field decomposes as
% $\hat{\mathcal{E}}(\mathbf{r},\omega) = \hat{\mathcal{E}}^P(\mathbf{r}) +
% \hat{\mathcal{E}}^B(\mathbf{r})$: a medium-assisted contribution driven by
% $\hat{\mathcal{P}}_N$ and a boundary-assisted contribution carrying incoming
% vacuum fluctuations $(\bar{\mathbf{n}}{\times})\hat{\mathcal{E}}_{\mathrm{in}}(s)$
% through $S$.}
%     \label{fig:quantum}
% \end{figure}
%\paragraph{Quantum Maxwell equation.}
Substituting the constitutive relation Eq.~\eqref{eq:constitutive_quantum} into the free-space Maxwell equation and absorbing the coherent response into the material tensor via $\bar{\varepsilon} = \mathbf{I} + \bar{\chi}$ gives the dressed Maxwell operator $\mathcal{M} = \mathcal{H} - k_0\bar{\varepsilon}$, and the field equation reduces to
\begin{equation}
\langle \mathbf r|\mathcal M|\hat{\mathcal E}\rangle
= k_0\,\hat{\mathcal{P}}_N(\mathbf{r},\omega).
\label{eq:quantum_maxwell}
\end{equation}

This is the same first-order system developed in Sec.~\ref{sec:green's-operator}, now driven by the Langevin noise, even in the absence of free currents in the medium, whose strength is fixed by the local dissipation through Eq.~\eqref{eq:FDT_dualP}. Inverting with the retarded Green's operator and applying Eq.~\eqref{eq:stratton_chu_intro} will split the solution into volume and boundary channels, each carrying an independent source of quantum fluctuations.

\paragraph{Field solution}
Eq.~\eqref{eq:quantum_maxwell} is a linear, operator-valued inhomogeneous equation whose solution is constructed with the retarded Green's operator $G$. Applying  Eq.~\eqref{eq:stratton_chu_intro} decomposes the total field into a medium-assisted (particular) and boundary-assisted (homogeneous) contribution,
\begin{equation}
\hat{\mathcal{E}}(\mathbf{r})
= \hat{\mathcal{E}}^P(\mathbf{r})
+ \hat{\mathcal{E}}^B(\mathbf{r}),
\label{eq:field_decomp}
\end{equation}
where
\begin{align}
\hat{\mathcal{E}}^P(\mathbf{r})
&= k_0\!\int g(\mathbf{r},\mathbf{r}')\,
\hat{\mathcal{P}}_N(\mathbf{r}')\,dV',
\label{eq:E_medium}\\
\hat{\mathcal{E}}^B(\mathbf{r})
&= -i\oint g(\mathbf{r},\mathbf s)\,
(\bar{n}\times)\,
\hat{\mathcal{E}}_{\mathrm{in}}(\mathbf s)\,dS.
\label{eq:E_boundary}
\end{align}
The medium-assisted field $\hat{\mathcal{E}}^P(\mathbf r)$ propagates Langevin noise from bulk absorption through the retarded Green's kernel; the boundary-assisted field $\hat{\mathcal{E}}^B$ carries incoming vacuum fluctuations from the surface $S$, the quantum counterpart of the classical input channel identified in Eq.~\eqref{eq:stratton_chu_intro}. The incoming tangential field $(\bar{n}\times)\hat{\mathcal{E}}_{\mathrm{in}}$ is quantized by imposing the equal-frequency commutator
\begin{multline}
\bigl[(\bar{n}\!\times\!\hat{\mathcal{E}}_{\mathrm{in}})_i(s,\omega),\,
(\bar{n}\!\times\!\hat{\mathcal{E}}_{\mathrm{in}})_j^\dagger(s',\omega')\bigr]\\
= -\frac{\hbar k_0}{2\pi \varepsilon_0}\,
(\bar{n}\!\times)_{ij}\,\delta_S(s\!-\!s')\,\delta(\omega\!-\!\omega'),
\label{eq:boundary_comm}
\end{multline}
the surface analogue of the volume fluctuation--dissipation relation Eq.~\eqref{eq:FDT_dualP}. The minus sign reflects the fact that $(\bar{n}\times)$ with outward normal is negative-definite on inward-propagating modes: incoming vacuum fluctuations carry inward Poynting flux, for which $\langle\hat{\Psi}|\bar{n}\times\hat{\Psi}\rangle_S < 0$. The volume noise strength is set by $\bar{\varepsilon}_I$, while the boundary noise strength is set by $(\bar{n}\times)$. The two noise sources are built from independent degrees of freedom; $\hat{\mathcal{P}}_N$ from the material bath operators $\hat{\mathbf{b}}(\mathbf{r},\omega)$ and $\hat{\mathcal{E}}_{\mathrm{in}}$ from the radiation modes on $S$; so all cross-commutators vanish and the total field commutator separates cleanly.
In geometries where the exterior region is itself structured, the boundary channels need not be free-space plane waves; they may belong to guided, radiative, or hybrid modes of an extended dielectric environment. The first-order Green-operator formalism accommodates this without modification, because the boundary contribution is expressed in terms of the tangential field trace rather than an assumed asymptotic basis.

%%%%%%%%%%%%%%%%%%%%%%%%%%%%%%%%%%%
\subsection{Field Commutation Relations}
\label{sec:commutators}
%%%%%%%%%%%%%%%%%%%%%%%%%%%%%%%%%%%
The commutator of Eq.~\ref{eq:field_decomp} then splits as
\begin{equation}
\bigl[\hat{\mathcal{E}},\,\hat{\mathcal{E}}^\dagger\bigr]
= \bigl[\hat{\mathcal{E}}^P,\,\hat{\mathcal{E}}^{P\dagger}\bigr]
+ \bigl[\hat{\mathcal{E}}^B,\,\hat{\mathcal{E}}^{B\dagger}\bigr].
\label{eq:commutator_split}
\end{equation}
We now verify that the quantized field Eq.~\eqref{eq:field_decomp} satisfies the correct equal-frequency commutation relations, and that the classical optical theorem of Sec.~\ref{sec:optical_theorem} is the identity that closes the quantum algebra.

%\paragraph{Volume contribution.}
Substituting $\hat{\mathcal{E}}^P = k_0\int g\,\hat{\mathcal{P}}_N\,dV'$ into the commutator and pulling the $c$-number Green's kernels outside gives
\begin{align}
\bigl[\hat{\mathcal{E}}^P_i(\mathbf{r}_1,\omega),\,
\hat{\mathcal{E}}^{P\dagger}_j(\mathbf{r}_2,\omega')\bigr]
= k_0^2\!\iint_{V\times V}
g_{i\alpha}(\mathbf{r}_1,\mathbf{r}')\nonumber\\
\bigl[\hat{\mathcal{P}}_{N,\alpha}(\mathbf{r}',\omega),\,
\hat{\mathcal{P}}_{N,\beta}^\dagger(\mathbf{r}'',\omega')\bigr]\,
g^\dagger_{\beta j}(\mathbf{r}'',\mathbf{r}_2)\,dV'\,dV''.
\label{eq:vol_step1}
\end{align}
Substituting the fluctuation--dissipation relation Eq.~\eqref{eq:FDT_dualP}, the spatial delta function collapses the double integral to a single integral and the frequency delta function imposes $\omega = \omega'$:
\begin{multline}
\bigl[\hat{\mathcal{E}}^P,\hat{\mathcal{E}}^{P\dagger}\bigr]
= \frac{\hbar k_0^2}{\pi \varepsilon_0}
\int g(\mathbf{r}_1,\mathbf{r}')\,
\bar{\varepsilon}_I(\mathbf{r}')\\
\times g^\dagger(\mathbf{r}',\mathbf{r}_2)\,dV'\;
\delta(\omega\!-\!\omega').
\label{eq:comm_vol}
\end{multline}

The same procedure applies to the boundary contribution.
Substituting the boundary propagation relation $\hat{\mathcal{E}}^B = \oint g\,(\bar{n}\times)\,\hat{\mathcal{E}}_{\mathrm{in}}\,dS$ and pulling the Green's kernels outside,
\begin{multline}
\bigl[\hat{\mathcal{E}}^B_i(\mathbf{r}_1,\omega),\,
\hat{\mathcal{E}}^{B\dagger}_j(\mathbf{r}_2,\omega')\bigr]
= \iint_{S\times S}
g_{i\kappa}(\mathbf{r}_1,s)\\
\times\bigl[(\bar{n}\times\hat{\mathcal{E}}_{\mathrm{in}})_\kappa(s),\,
(\bar{n}\times\hat{\mathcal{E}}_{\mathrm{in}}^\dagger)_\ell(s')\bigr]\,
g^\dagger_{\ell j}(s',\mathbf{r}_2)\,dS\,dS'.
\label{eq:bdy_step1}
\end{multline}
Substituting the boundary commutator Eq.~\eqref{eq:boundary_comm}, the surface delta function collapses the double integral in the same way the spatial delta collapsed the volume integral,
\begin{multline}
\bigl[\hat{\mathcal{E}}^B,\hat{\mathcal{E}}^{B\dagger}\bigr]
= -\frac{\hbar k_0}{2\pi \varepsilon_0}
\oint g(\mathbf{r}_1,s)\,
(\bar{n}\!\times)\\
\times g^\dagger(s,\mathbf{r}_2)\,dS\;
\delta(\omega\!-\!\omega').
\label{eq:comm_bdy}
\end{multline}
Unlike the volume term, this contribution survives in a lossless medium: it encodes the rate at which vacuum fluctuations flow through $S$, the quantum counterpart of the classical surface Poynting flux.

The total commutator is the sum of the volume and boundary contributions.
Adding Eq.~\eqref{eq:comm_vol} and Eq.~\eqref{eq:comm_bdy},
\begin{multline}
\bigl[\hat{\mathcal{E}}_i(\mathbf{r}_1,\omega),\,
\hat{\mathcal{E}}_j^\dagger(\mathbf{r}_2,\omega')\bigr]
= \frac{\hbar k_0}{\pi \varepsilon_0}\biggl[
 k_0\!\int g\,\bar{\varepsilon}_I\,g^\dagger\,dV'\\
- \frac{1}{2}\oint g\,(\bar{n}\times)\,g^\dagger\,dS
\biggr]\delta(\omega-\omega').
\label{eq:comm_sum}
\end{multline}
The bracketed expression is precisely the right-hand side of the generalized optical theorem Eq.~\eqref{eq:optical_theorem_kernel}, which can be written equivalently as  
\begin{equation}
\begin{split}
&\big(g(\mathbf{r}_1,\mathbf{r}_2)
- g^\dagger(\mathbf{r}_1,\mathbf{r}_2)\big)/2i \\
&= k_0\!\int g(\mathbf{r}_1,\mathbf{r}')\,
\bar{\varepsilon}_I(\mathbf{r}')\,g^\dagger(\mathbf{r}',\mathbf{r}_2)\,dV'\\
&\quad - \frac{1}{2}\oint g(\mathbf{r}_1,s)\,
(\bar{n} \times)\,g^\dagger(s,\mathbf{r}_2)\,dS,
\end{split}
\label{eq:OT_recall}
\end{equation}
where we have used the reciprocity relation to write both sides with the argument ordering $(\mathbf{r}_1,\mathbf{r}_2)$ matching the commutator. Substituting Eq.~\eqref{eq:OT_recall} into Eq.~\eqref{eq:comm_sum} and interpreting $(\bar{n}\times)$ with inward normal yields the final result,
\begin{multline}
\bigl[\hat{\mathcal{E}}_i(\mathbf{r}_1,\omega),\,
\hat{\mathcal{E}}_j^\dagger(\mathbf{r}_2,\omega')\bigr]\\
= \frac{\hbar k_0}{\pi \varepsilon_0}\,
\mathrm{Im}\,g_{ij}(\mathbf{r}_1,\mathbf{r}_2,\omega)\,
\delta(\omega\!-\!\omega'),
\label{eq:commutator_final}
\end{multline}
where $\mathrm{Im}\,g (\mathbf r_1, \mathbf r_2)\equiv (g(\mathbf r_1, \mathbf r_2) - g^\dagger(\mathbf r_1, \mathbf r_2))/(2i)$ denotes the anti-Hermitian part of the retarded Green's kernel under the energy adjoint of Sec.~\ref{sec:greens_adjoint}. The cancellation is not accidental: the fluctuation--dissipation theorem fixes the strength of each noise channel to match the dissipation it compensates, and the optical theorem is the identity stating that these channels partition the total loss in the correct proportion. The commutator is thus determined entirely by the single classical object $\mathrm{Im}\,g$, with no separate reference to bulk absorption or surface flux; the same first-order Green's operator that propagates classical fields also determines the quantum fluctuation structure, with the optical theorem serving as the bridge between the two.

%%%%%%%%%%%%%%%%%%%%%%%%%%%%%%%%%%%
\section{Quantum Transfer and Input--Output Relations}
\label{sec:quantum_io}
%%%%%%%%%%%%%%%%%%%%%%%%%%%%%%%%%%%
%\jon{i will check this. this probably has errors right now.}
The classical transfer operators of Sec.~\ref{sec:transfer} fit directly to the quantum theory.  Let $V_{21}$ be the region between two nonintersecting surfaces $S_1$ and $S_2$, and assume that no free currents are present in $V_{21}$.  We define the quantum tangential dual trace on each surface by
\begin{equation}
\hat{\Psi}_i(\mathbf s,\omega)
\equiv
(\bar{n}_i\times)\hat{\mathcal E}(\mathbf s,\omega),
\qquad \mathbf s\in S_i.
\label{eq:psi_quantum_def}
\end{equation}
Applying the tangential trace $(\bar{n}_2\times)$ to Eq.~\eqref{eq:field_decomp} gives the exact surface-to-surface quantum transfer law
\begin{multline}
\hat{\Psi}_2(\mathbf s_2,\omega)
=
\oint_{S_1}\!
T_{21}(\mathbf s_2,\mathbf s_1,\omega)\,
\hat{\Psi}_1(\mathbf s_1,\omega)\,dS_1\\
+\hat{\Psi}_{N,21}(\mathbf s_2,\omega),
\label{eq:quantum_transfer}
\end{multline}
where
\begin{equation}
T_{21}(\mathbf s_2,\mathbf s_1,\omega)
=
-i(\bar{n}_2\times)\,
g(\mathbf s_2,\mathbf s_1,\omega)
\label{eq:quantum_transfer_kernel}
\end{equation}
is the same retarded transfer kernel as in the classical theory, and
\begin{multline}
\hat{\Psi}_{N,21}(\mathbf s_2,\omega)
=
k_0\!\int_{V_{21}}\!
(\bar{n}_2\!\times)\\
\times g(\mathbf s_2,\mathbf r,\omega)\,
\hat{\mathcal P}_N(\mathbf r,\omega)\,dV
\label{eq:transfer_noise}
\end{multline}
is the noise added by the absorptive medium between the two surfaces.
Equation~\eqref{eq:quantum_transfer} is the exact first-order quantum input--output relation: the outgoing boundary state on $S_2$ is the sum of a coherently propagated input from $S_1$ and a Langevin noise contribution generated in the intervening medium.  Since the incoming boundary operators and the material bath operators are independent, their cross-commutators vanish, and the output commutator takes the form
\begin{equation}
[\hat{\Psi}_2,\hat{\Psi}_2^\dagger]
=
T_{21}\,[\hat{\Psi}_1,\hat{\Psi}_1^\dagger]\,T_{21}^\dagger
+
[\hat{\Psi}_{N,21},\hat{\Psi}_{N,21}^\dagger].
\label{eq:io_comm_split}
\end{equation}
If the input trace on $S_1$ satisfies the canonical boundary commutator
\begin{multline}
\bigl[\hat{\Psi}_{1,i}(\mathbf s,\omega),\,
\hat{\Psi}_{1,j}^\dagger(\mathbf s',\omega')\bigr]\\
=
-\frac{\hbar k_0}{2\pi\varepsilon_0}\,
(\bar{n}_1\!\times)_{ij}\,
\delta_S(\mathbf s\!-\!\mathbf s')\,\delta(\omega\!-\!\omega'),
\label{eq:psi_in_comm}
\end{multline}
then the same must hold for the output trace on $S_2$:
\begin{multline}
\bigl[\hat{\Psi}_{2,i}(\mathbf s,\omega),\,
\hat{\Psi}_{2,j}^\dagger(\mathbf s',\omega')\bigr]\\
=
-\frac{\hbar k_0}{2\pi\varepsilon_0}\,
(\bar{n}_2\!\times)_{ij}\,
\delta_S(\mathbf s\!-\!\mathbf s')\,\delta(\omega\!-\!\omega').
\label{eq:psi_out_comm}
\end{multline}
Comparing Eqs.~\eqref{eq:io_comm_split}--\eqref{eq:psi_out_comm}, the noise commutator is fixed to be
\begin{equation}
\begin{split}
&\bigl[\hat{\Psi}_{N,21}(\mathbf s_2,\omega),\,
\hat{\Psi}_{N,21}^\dagger(\mathbf s_2',\omega')\bigr]
\\
&\quad=
-\frac{\hbar k_0}{2\pi\varepsilon_0}\delta(\omega\!-\!\omega')
\biggl[
(\bar{n}_2\!\times)\,\delta_S(\mathbf s_2\!-\!\mathbf s_2')\\
&\qquad-\!
\oint_{S_1}\!
T_{21}(\mathbf s_2,\mathbf s_1)\,
(\bar{n}_1\!\times)
T_{21}^\dagger(\mathbf s_1,\mathbf s_2')\,dS_1
\biggr].
\end{split}
\label{eq:transfer_noise_comm}
\end{equation}

% From Eq.~\eqref{eq:transfer_noise} and using Eq. \eqref{eq:FDT_dualP}
% \begin{equation}
% \begin{split}
% &\bigl[\hat{\Psi}_{N,21}(\mathbf s_2,\omega),\,
% \hat{\Psi}_{N,21}^\dagger(\mathbf s_2',\omega')\bigr]
% \\
% &\quad=
% \frac{\hbar k_0^2}{\pi\varepsilon_0} \int_V 
% (\bar{n}_2\!\times)\,g(\mathbf s_2,\mathbf r)\bar{\varepsilon}_I(\mathbf r,\omega)\\
% &\qquad\qquad\qquad\quad\times g^\dagger(\mathbf{r}, \mathbf{s}_2')(\bar{n}_2\!\times)^\dagger
% \delta(\omega\!-\!\omega').
% \end{split}
% % \label{eq:transfer_noise_comm}
% \end{equation}
Thus the amount of quantum noise added by the medium is exactly the amount required to preserve the canonical boundary commutator after propagation.
Equation~\eqref{eq:transfer_noise_comm} is the input--output analogue of the generalized optical theorem: coherent transfer and added noise are not independent, but are constrained by the same classical Green operator.
In the lossless case, $\bar{\varepsilon}_I=0$ and the bulk Langevin noise vanishes, so $\hat{\Psi}_{N,21}=0$.  Equation~\eqref{eq:transfer_noise_comm} then reduces to the pseudo-unitarity condition
\begin{multline}
\oint_{S_1}\!
T_{21}(\mathbf s_2,\mathbf s_1)\,
(\bar{n}_1\!\times)\,
T_{21}^\dagger(\mathbf s_1,\mathbf s_2')\,dS_1\\
=
(\bar{n}_2\!\times)\,\delta_S(\mathbf s_2\!-\!\mathbf s_2'),
\label{eq:transfer_pseudounitary}
\end{multline}
showing that a lossless transfer operator preserves the surface symplectic metric exactly.  Equation~\eqref{eq:quantum_transfer} generalizes the beam-splitter relation of quantum optics to a spatially resolved, frequency-dependent setting in which the medium itself plays the role of the beam splitter and the distributed Langevin noise replaces the vacuum port. In a translationally invariant planar geometry, the surface integrals reduce to matrix multiplication at fixed transverse wavevector, and Eqs.~\eqref{eq:quantum_transfer}--\eqref{eq:transfer_pseudounitary} become the usual quantum transfer-matrix or scattering-matrix relations of waveguide and cavity quantum optics.

%%%%%%%%%%%%%%%%%%%%%%%%%%%%%%%%%%%%%%%%%%%%%%%%%%%%%%%%%%%%
\section{Conclusion}

We have presented a first-order operator formalism for macroscopic QED that retains both $\mathbf{E}$ and $\mathbf{H}$ in a single dual field and keeps the open-system boundary terms throughout. The adjoint structure of the Maxwell operator under two inner products yields the major classical identities, power balance, Lorentz reciprocity, the Stratton--Chu representation, and a generalized optical theorem, from symmetries rather than vector calculus.

The generalized optical theorem plays a dual role: classically it decomposes dissipation into bulk absorption and boundary radiation, and quantum-mechanically it is the identity that closes the field commutator as $[\hat{\mathcal{E}},\hat{\mathcal{E}}^\dagger] \propto \mathrm{Im}\,g$. The entire quantum fluctuation structure is determined by a single classical object, $\mathrm{Im}\,g$, with the two dissipation channels mapping directly onto the two quantum noise sources: Langevin noise from material absorption and vacuum fluctuations incoming at the boundary. The interior quantum field is compactly expressed as $\hat{\mathcal{E}} = k_0\int g\,\hat{\mathcal{P}}_N\,dV' - i\oint g\,(\bar{n}\times)\hat{\mathcal{E}}_\mathrm{in}\,dS$, making the two noise contributions and the role of the Green operator explicit.

A key feature is that the boundary channels are expressed in terms of tangential field traces, requiring no assumption about the exterior environment. This resolves the known incompleteness of volume-only quantization schemes for finite objects~\cite{drezet2017quantizing,ciattoni20240719quantum,stefano2001mode} and, unlike previous extensions that rely on free-space plane-wave asymptotics, applies equally when structured dielectric media extend to the boundary, as in waveguide input--output problems.

Because the formalism is built on the retarded Green operator rather than a modal decomposition~\cite{kristensen2014quasinormal}, it applies to any electromagnetic environment whose Green's function can be obtained, numerically or analytically. This extends macroscopic QED to the growing class of nanophotonic systems~\cite{molesky2018inverse} that lie beyond cavity and waveguide models: inverse-designed devices, photonic-crystal bandgaps~\cite{hood2016atom}, hybrid cavity QED with molecular defects~\cite{lange2026hybrid}, and superradiant atom arrays~\cite{lange2024superradiant,asenjo-garcia2017exponential}. In each case the quantization procedure is unchanged; only the Green's function input differs.

The transfer operators compose under cascading: chaining $N$ photonic elements yields $T = T_N \cdots T_1$, with each stage adding exactly the Langevin noise required by the optical theorem to preserve the canonical commutator at the output. This makes the formalism a natural framework for cascaded quantum networks built from heterogeneous photonic components, interfacing directly with standard numerical electromagnetic solvers that already compute the retarded Green's function for complex geometries.

Finally, because the first-order formalism treats $\mathbf{E}$ and $Z_0\mathbf{H}$ on equal footing, it extends naturally to bianisotropic media with magnetoelectric coupling~\cite{scheel2009duality,scheel2012macroscopic,rapp2025purcell}, where the second-order Helmholtz equation for $\mathbf{E}$ alone is no longer sufficient. This would bring chiral, magneto-optic, nonreciprocal, and topological photonic systems~\cite{silveirinha2018topological} within the same quantum input--output framework. A modal input--output theory, in which the boundary fields are decomposed into guided and radiation modes of the input and output waveguides, will follow in subsequent work.
\section{Acknowledgments}

This work is supported by an Energy Frontier Research Center funded by the U.S. Department of Energy (DOE), Office of Science, Basic Energy Sciences (BES), under award DE-SC0025620. We thank Stuart Masson for a careful reading of the manuscript and for valuable feedback.
%%%%%%%%%%%%%%%%%%%%%%%%%%%%%%%%%%%%%%%%%%%%%%%%
% \newpage
\appendix
\input{appendix}
\bibliography{references}

\end{document}

%% file: appendix.tex
% \onecolumngrid
%%%%%%%%%%%%%%%%%%%%%%%%%%%%%%%%%%%%%%%%%%%%%%%%%
\newpage

\section{Vector Green's theorem for the dual curl}
\label{app:vector_green}
We derive the adjoint identity for the dual curl operator $\bar{\nabla}\times$ used in Sec. VI. The goal is to evaluate $\langle\mathcal{E}_1|\bar{\nabla}\times\mathcal{E}_2\rangle$ and express it in terms of $\bar{\nabla}\times\mathcal{E}_1$ and a surface integral over the boundary $S$ of $V$.

Expanding using $\bar{\nabla}\times\mathcal{E} = [\nabla\times(Z_0\mathbf{H}),\,-\nabla\times\mathbf{E}]^T$, the integrand is
\begin{equation}
\begin{split}
\mathcal{E}_1^\dagger (\bar{\nabla} \times \mathcal{E}_2) ={} & \mathbf{E}_1^* \cdot \nabla \times (Z_0 \mathbf{H}_2) \\
& - (Z_0 \mathbf{H}_1)^* \cdot (\nabla \times \mathbf{E}_2).
\end{split}
\end{equation}

Applying the identity $\mathbf{A}\cdot(\nabla\times\mathbf{B}) = \mathbf{B}\cdot(\nabla\times\mathbf{A}) + \nabla\cdot(\mathbf{B}\times\mathbf{A})$ to each term and using $\nabla\times\mathbf{F}^* = (\nabla\times\mathbf{F})^*$, one finds
\begin{equation}
\begin{split}\mathcal{E}_1^\dagger\,(\bar{\nabla}\times\mathcal{E}_2) = -(\bar{\nabla}\times\mathcal{E}_1)^\dagger\cdot\mathcal{E}_2\\ + \nabla\cdot\bigl((Z_0\mathbf{H}_2)\times\mathbf{E}_1^* - \mathbf{E}_2\times(Z_0\mathbf{H}_1^*)\bigr),\end{split} \end{equation}

where the first term is verified by noting $(\bar{\nabla}\times\mathcal{E}_1)^\dagger = [(\nabla\times(Z_0\mathbf{H}_1))^*,\,-(\nabla\times\mathbf{E}_1)^*]$. Integrating over $V$ and applying the divergence theorem,

\begin{equation*}\begin{split}\int_V \mathcal{E}_1^\dagger\,(\bar{\nabla}\times\mathcal{E}_2)\,dV = -\int_V (\bar{\nabla}\times\mathcal{E}_1)^\dagger\cdot\mathcal{E}_2\,dV\\
+ \oint_S \hat{\mathbf{n}}\cdot\bigl((Z_0\mathbf{H}_2)\times\mathbf{E}_1^* - \mathbf{E}_2\times(Z_0\mathbf{H}_1^*)\bigr)\,dS.
\end{split}\end{equation*}

The surface integrand is rewritten using $\hat{\mathbf{n}}\cdot(\mathbf{A}\times\mathbf{B}) = \mathbf{B}\cdot(\hat{\mathbf{n}}\times\mathbf{A})$:

\begin{align*}
\begin{split}\hat{\mathbf{n}}\cdot\bigl((Z_0\mathbf{H}_2)\times\mathbf{E}_1^* - \mathbf{E}_2\times(Z_0\mathbf{H}_1^*)\bigr) = \\
\mathbf{E}_1^*\cdot(\hat{\mathbf{n}}\times Z_0\mathbf{H}_2) - (Z_0\mathbf{H}_1)^*\cdot(\hat{\mathbf{n}}\times\mathbf{E}_2) = \\ \mathcal{E}_1^\dagger\cdot(\bar{n}\times\mathcal{E}_2),
\end{split}
\end{align*}

where the last equality defines the dual surface operator
$$\bar{n}\times\mathcal{E}(\mathbf r) \equiv \begin{bmatrix}\hat{\mathbf{n}}\times(Z_0\mathbf{H}(\mathbf r)) \\ -\hat{\mathbf{n}}\times\mathbf{E}(\mathbf r)\end{bmatrix}.$$

This operator extracts the tangential field components with the same symplectic sign structure as $\bar{\nabla}\times$. Collecting terms and introducing the surface inner product $\langle\mathcal{E}_1|\bar{n}\times\mathcal{E}_2\rangle_S \equiv \oint_S \mathcal{E}_1^\dagger\cdot(\bar{n}\times\mathcal{E}_2)\,dS$, we arrive at the Green's theorem for the dual curl:
\begin{align}\langle\mathcal{E}_1|\bar{\nabla}\times\mathcal{E}_2\rangle = -\langle\bar{\nabla}\times\mathcal{E}_1|\mathcal{E}_2\rangle + \langle\mathcal{E}_1|\bar{n}\times\mathcal{E}_2\rangle_S.\end{align}

The dual curl is formally anti-self-adjoint, with the boundary form given by the surface inner product. Multiplying by $i$, the Maxwell Hamiltonian $\mathcal{H} = i\bar{\nabla}\times$ satisfies
\begin{align}\langle\mathcal{E}_1|\mathcal{H}\mathcal{E}_2\rangle = \langle\mathcal{H}\mathcal{E}_1|\mathcal{E}_2\rangle + i\langle\mathcal{E}_1|\bar{n}\times\mathcal{E}_2\rangle_S,\end{align}

confirming that $\mathcal{H}$ is formally self-adjoint with boundary form $\mathcal{B} = i\langle\mathcal{E}_1|\bar{n}\times\mathcal{E}_2\rangle_S$. This is the only integration-by-parts identity needed in the paper; all subsequent results follow from operator algebra.

%%%%%%%%%%%%%%%%%%%%%%%%%%%%%%%%%%%%%%%%%%%%%%%%%
\section{Reciprocal adjoint derivations}
\label{app:reciprocal_adjoint}
%%%%%%%%%%%%%%%%%%%%%%%%%%%%%%%%%%%%%%%%%%%%%%%%%
This appendix derives the reciprocal identities stated in Sec.~\ref{sec:reciprocity}. We introduce the reciprocal adjoint $\mathcal{O}^\sharp$ of an operator $\mathcal{O}$, defined with respect to the reciprocal pairing $\langle\mathcal{E}_1|\mathcal{E}_2\rangle_R = \int_V \mathcal{E}_1^T\,\Pi\,\mathcal{E}_2\,dV$ (Eq.~\eqref{eq:reciprocal_ip}) by
\begin{equation}
  \langle\mathcal{E}_1|\mathcal{O}\mathcal{E}_2\rangle_R
  = \langle\mathcal{O}^\sharp\mathcal{E}_1|
    \mathcal{E}_2\rangle_R
  + \mathcal{B}_R[\mathcal{E}_1,\mathcal{E}_2],
  \label{eq:app_sharp_def}
\end{equation}
where $\mathcal{B}_R$ is a boundary form that vanishes when either field has compact support in $V$. Because the pairing is bilinear rather than sesquilinear, the reciprocal adjoint involves the transpose of multiplication operators rather than the Hermitian conjugate.

\subsection{Reciprocal adjoint of the Maxwell operator}
\label{app:M_sharp}

We compute $\mathcal{M}^\sharp$ by treating the differential and material parts of $\mathcal{M}=\mathcal{H}-k_0\bar{\varepsilon}$ separately.

The differential part $\mathcal{H} = i\bar{\nabla}\times$ is moved across the pairing by the same integration by parts used for the $L^2$ adjoint (Appendix~\ref{app:vector_green}). The curl identity is insensitive to whether the outer product involves conjugation or transposition, so $\mathcal{H}^\sharp = \mathcal{H}$ with the same boundary form:
\begin{equation}
  \langle\mathcal{E}_1|\mathcal{H}\mathcal{E}_2\rangle_R
  = \langle\mathcal{H}\mathcal{E}_1|\mathcal{E}_2\rangle_R
  + i\langle\mathcal{E}_1|\bar{n}\times\mathcal{E}_2\rangle_{R,S},
  \label{eq:app_H_IBP}
\end{equation}
where $\langle\mathcal{E}_1|\bar{n}\times\mathcal{E}_2\rangle_{R,S} \equiv \oint_S \mathcal{E}_1^T\,\Pi\,(\bar{n}\times\mathcal{E}_2)\,dS$.

The material part $\bar{\varepsilon}$ acts by pointwise multiplication, so no integration by parts is needed. Since $\Pi$ and $\bar{\varepsilon} = \mathrm{diag}(\boldsymbol{\varepsilon},\boldsymbol{\mu})$ are both block diagonal they commute, and transposing $\mathcal{E}_1^T(\Pi\bar{\varepsilon})$ under the integral gives $(\Pi\bar{\varepsilon})^T = \bar{\varepsilon}^T\Pi$, so
\begin{equation}
  \bar{\varepsilon}^\sharp = \bar{\varepsilon}^T.
  \label{eq:app_epssharp}
\end{equation}
Under the $L^2$ pairing the analogous result is $\bar{\varepsilon}^\dagger$; here, the absence of complex conjugation exposes the transpose instead.

Combining these,
\begin{equation}
  \mathcal{M}^\sharp = \mathcal{H} - k_0\bar{\varepsilon}^T.
  \label{eq:app_Msharp}
\end{equation}
For a reciprocal medium ($\bar{\varepsilon}^T = \bar{\varepsilon}$), the Maxwell operator is self-adjoint under the reciprocal pairing: $\mathcal{M}^\sharp = \mathcal{M}$. The departure from self-adjointness is
\begin{equation}
  \mathcal{M}^\sharp - \mathcal{M} = k_0(\bar{\varepsilon} - \bar{\varepsilon}^T),
  \label{eq:app_Msharp_departure}
\end{equation}
which vanishes for any medium with symmetric $\boldsymbol{\varepsilon}$ and $\boldsymbol{\mu}$ tensors, including lossy, dispersive, and inhomogeneous media. This contrasts with the energy-adjoint structure, where the departure is controlled by $\bar{\varepsilon}_I = (\bar{\varepsilon}-\bar{\varepsilon}^\dagger)/2i$ and vanishes only for lossless media. Reciprocity is a statement about source--observer interchange and is compatible with arbitrary loss; the energy adjoint probes power balance and is sensitive to absorption.

\subsection{Reciprocal adjoint identity and Lorentz surface form}
\label{app:reciprocal_identity}

Substituting $\mathcal{M}^\sharp = \mathcal{M}$ into Eq.~\eqref{eq:app_sharp_def} gives the reciprocal adjoint identity stated in the main text (Eq.~\eqref{eq:reciprocal_adjoint_identity}):
\begin{equation}
  \langle\mathcal{E}_1|\mathcal{M}\mathcal{E}_2\rangle_R
  - \langle\mathcal{M}\mathcal{E}_1|\mathcal{E}_2\rangle_R
  = i\langle\mathcal{E}_1|\bar{n}\times\mathcal{E}_2\rangle_{R,S}.
  \label{eq:reciprocal_adjoint_identity_recip}
\end{equation}
Unlike the energy-adjoint identity, there is no volume term: material loss does not spoil reciprocity. The surface integrand expands via the scalar triple-product identity as
\begin{align}
  \mathcal{E}_1^T\,\Pi\,(\bar{n}\times\mathcal{E}_2)
  &= \mathbf{E}_1\cdot(\hat{\mathbf{n}}\times Z_0\mathbf{H}_2)
     + \\&\qquad Z_0\mathbf{H}_1\cdot(\hat{\mathbf{n}}\times\mathbf{E}_2)
  \nonumber\\
  &= Z_0\,\hat{\mathbf{n}}\cdot\bigl(
     \mathbf{E}_2\times\mathbf{H}_1
     - \mathbf{E}_1\times\mathbf{H}_2\bigr).
  \label{eq:lorentz_surface_expand}
\end{align}
This Lorentz surface form plays the same structural role as the Poynting flux in the energy-adjoint identity, but compares two distinct field states rather than pairing a field with its complex conjugate. The Lorentz reciprocity theorem (Eq.~\eqref{eq:lorentz_reciprocity}) follows immediately by substituting $\mathcal{M}|\mathcal{E}_{1,2}\rangle = |\mathcal{S}_{1,2}\rangle$.

\subsection{Reciprocal adjoint of the Green operator}
\label{sec:greens_reciprocal_adjoint}

The reciprocal adjoint of the integral operator $G$ is defined by $\langle\mathcal{S}_1|G\mathcal{S}_2\rangle_R = \langle G^\sharp\mathcal{S}_1|\mathcal{S}_2\rangle_R$. Since $G$ acts by integration rather than differentiation, no boundary term is generated (the same reasoning as for the $L^2$ adjoint in Sec.~\ref{sec:greens_adjoint}). Expanding in kernel form, interchanging the order of integration, and comparing sides yields
\begin{equation}
  g^\sharp(\mathbf{r},\mathbf{r}') = \Pi\,g^T(\mathbf{r}',\mathbf{r})\,\Pi.
  \label{eq:gsharp_kernel_recip}
\end{equation}
For a reciprocal medium, $\mathcal{M}^\sharp = \mathcal{M}$ implies $G^\sharp = G$, recovering the kernel symmetry stated in the main text (Eq.~\eqref{eq:green_reciprocity}): $g(\mathbf{r},\mathbf{r}') = \Pi\,g^T(\mathbf{r}',\mathbf{r})\,\Pi$. Unlike the energy adjoint, no complex conjugation appears, so the reciprocal adjoint preserves the causal character of the propagator: a retarded Green operator remains retarded.

\section{Quantization in Dispersive-Lossy Media}
\label{app:quantization}
%------------------------
%------------------------------------------------------------------
The dispersive, absorptive medium is modeled as two independent
continua of harmonically bound oscillators: a polarization density
$\hat{\mathbf{P}}(\mathbf{r})$ carrying the electric response, and a
magnetization density $\hat{\mathbf{M}}(\mathbf{r})$ carrying the
magnetic response.  Each represents the collective displacement of
bound charges or magnetic dipoles per unit volume, and each is
coupled to its own local bosonic reservoir responsible for
irreversible damping. We introduce the
dual polarization vector
\begin{equation}
\hat{\mathcal{P}}(\mathbf{r})
\equiv
\begin{pmatrix}
  \hat{\mathbf{P}}(\mathbf{r})/\varepsilon_0 \\[4pt]
  Z_0\hat{\mathbf{M}}(\mathbf{r})
\end{pmatrix}.
\label{eq:dualP}
\end{equation}
The canonical conjugate momentum to
$\hat{\mathcal{P}}$ is defined so that the standard commutation
relation holds identically in both sectors:
\begin{equation}
\hat{\boldsymbol{\Pi}}_{\mathcal{P}}(\mathbf{r})
\equiv
\begin{pmatrix}
  \varepsilon_0\,\hat{\boldsymbol{\Pi}}_P(\mathbf{r}) \\[4pt]
  \hat{\boldsymbol{\Pi}}_M(\mathbf{r})/Z_0
\end{pmatrix},
\label{eq:dualPi}
\end{equation}
where $\hat{\boldsymbol{\Pi}}_P = M_P\dot{\hat{\mathbf{P}}}$ is the canonical
momentum of the electric polarization (Eq.~\eqref{eq:dualP}) and
$\hat{\boldsymbol{\Pi}}_M = M_M\dot{\hat{\mathbf{M}}}$ its magnetic
counterpart.  The rescaling factors $\varepsilon_0$ and $1/Z_0$ are
chosen so that the commutation relation
\begin{equation}
\bigl[\hat{\mathcal{P}}_i(\mathbf{r}),\,
      \hat{\Pi}_{\mathcal{P},j}(\mathbf{r}')\bigr]
= i\hbar\,\delta_{ij}\,\delta^{(3)}(\mathbf{r}-\mathbf{r}')
\label{eq:dualPPi_comm}
\end{equation}
holds for both $i,j \in \{1,2,3\}$ (electric sector) and
$i,j \in \{4,5,6\}$ (magnetic sector) without additional prefactors.
Cross-sector commutators vanish,
$[\hat{\mathcal{P}}_{1,i},\hat{\Pi}_{\mathcal{P}_2,j}] = 0$,
because the electric and magnetic oscillators are physically
independent.

Substituting the rescalings~\eqref{eq:dualP}--\eqref{eq:dualPi}
into the individual kinetic energies,
$\hat{\Pi}_P^2/(2M_P)$ and $\hat{\Pi}_M^2/(2M_M)$, yields a common
form controlled by the $6\times 6$ block-diagonal
 effective mass matrix
\begin{equation}
\bar{M}(\mathbf{r})
\equiv
\begin{pmatrix}
  \widetilde{M}_e(\mathbf{r})\,I_3 & 0 \\[4pt]
  0 & \widetilde{M}_m(\mathbf{r})\,I_3
\end{pmatrix},
\label{eq:Mbar}
\end{equation}
where the two blocks are
\begin{equation}
\widetilde{M}_e(\mathbf{r}) \equiv M_P\,\varepsilon_0^2,
\qquad
\widetilde{M}_m \equiv \frac{M_M}{Z_0^2}
= \frac{M_M\,\varepsilon_0}{\mu_0}.
\label{eq:effectivemasses}
\end{equation}
Here $M_P = m_e/[N(\mathbf{r})e^2]$ is the electric effective mass and $M_M$ is its magnetic analogue, defined by
$\hat{\boldsymbol{\Pi}}_M = M_M\dot{\hat{\mathbf{M}}}$.  
The dual resonance-frequency and damping matrices are likewise
block diagonal, $\bar{\Omega}_0^2(\mathbf{r})=\text{diag}\{\omega_{0e}^2(\mathbf{r})\,I_3, \;\omega_{0m}^2(\mathbf{r})\,I_3\}$
and $\bar{\Gamma}(\mathbf{r})=\text{diag}\{\gamma_e(\mathbf{r})\,I_3, \,\gamma_m(\mathbf{r})\,I_3\}$
where $\omega_{0e},\omega_{0m}$ are the bare resonance frequencies
and $\gamma_e,\gamma_m$ the Markovian damping rates of the
electric and magnetic Lorentz oscillators, respectively.

Each spatial point supports two independent bosonic reservoirs, one
for each sector.  We combine them into the dual bath operator
\begin{equation}
\hat{\mathbf{b}}(\mathbf{r},\omega)
\equiv
\begin{pmatrix}
  \hat{\mathbf{b}}_e(\mathbf{r},\omega) \\[4pt]
  \hat{\mathbf{b}}_m(\mathbf{r},\omega)
\end{pmatrix},
\label{eq:dualb}
\end{equation}
where $\hat{\mathbf{b}}_e$ and $\hat{\mathbf{b}}_m$ are the
3-component electric and magnetic reservoir fields.  All six
components satisfy bosonic commutation relations,
\begin{equation}
\bigl[\hat{b}_\alpha(\mathbf{r},\omega),\,
      \hat{b}_\beta^\dagger(\mathbf{r}',\omega')\bigr]
= \delta_{\alpha\beta}\,
  \delta^{(3)}(\mathbf{r}-\mathbf{r}')\,
  \delta(\omega-\omega'),
\label{eq:dualb_comm}
\end{equation}
for $\alpha,\beta = 1,\ldots,6$
with the cross-sector commutators
$[\hat{b}_{e,i},\hat{b}_{m,j}^\dagger] = 0$ following from
the physical independence of the two reservoirs.  As in the
electric-only case, the dual bath and the electromagnetic field
are never coupled directly; all dissipation is mediated by
the dual polarization $\hat{\mathcal{P}}$.

The impedance normalization~\eqref{eq:dualP}--\eqref{eq:dualPi}
shifts the effective coupling of each sector to its reservoir.
Tracing through the rescalings from the physical couplings
$\kappa_e(\mathbf{r},\omega)$ and $\kappa_m(\mathbf{r},\omega)$
(defined in analogy with the scalar $\kappa$ of
Eq.~\eqref{eq:H_total}) yields the dual coupling spectrum
\begin{equation}
\bar{\kappa}(\mathbf{r},\omega)
\equiv
\begin{pmatrix}
  \varepsilon_0\,\kappa_e(\mathbf{r},\omega)\,I_3 & 0 \\[4pt]
  0 & \kappa_m(\mathbf{r},\omega)/Z_0\,I_3
\end{pmatrix}.
\label{eq:kappa_dual}
\end{equation}
% The electric block picks up a factor of $\varepsilon_0$ because
% $\hat{P} = \varepsilon_0\hat{\mathcal{P}}_1$, and the magnetic block
% a factor of $1/Z_0$ because $\hat{M} = \hat{\mathcal{P}}_2/Z_0$.
Off-diagonal blocks of $\bar{\kappa}$ vanish since the electric
oscillators couple only to $\hat{\mathbf{b}}_e$ and the magnetic
oscillators only to $\hat{\mathbf{b}}_m$; a bianisotropic medium
with magnetoelectric coupling would populate these off-diagonal
blocks at this level.

%------------------------------------------------------------------
\subsection{Dual macroscopic Hamiltonian.}
%------------------------------------------------------------------
Collecting all five contributions, the full dual macroscopic
field-matter-bath Hamiltonian is
\begin{equation}
\hat{H}
= \hat{H}_{\mathrm{EM}}
+ \hat{H}_{\mathcal{P}}
+ \hat{H}_{\mathcal{P}\text{-}\mathcal{E}}
+ \hat{H}_{\mathrm{bath}}
+ \hat{H}_{\mathcal{P}\text{-bath}},
\label{eq:H_dual_total}
\end{equation}
with each term given explicitly as follows.

The free-field energy is written in terms of the dual field using
$\mu_0 = \varepsilon_0 Z_0^2$:
\begin{equation}
\hat{H}_{\mathrm{EM}}
= \frac{\varepsilon_0}{2}\int
  \bigl\lvert\hat{\mathcal{E}}(\mathbf{r})\bigr\rvert^2
  \,d^3r
\label{eq:H_EM_dual}
\end{equation}
The kinetic and potential energies of the dual polarization
oscillators are
\begin{align}
\hat{H}_{\mathcal{P}}
&= \frac{1}{2}\int
    \hat{\boldsymbol{\Pi}}_{\mathcal{P}}^\dagger(\mathbf{r})\;
    \bar{M}^{-1}(\mathbf{r})\;
    \hat{\boldsymbol{\Pi}}_{\mathcal{P}}(\mathbf{r})
  \,d^3r
\nonumber\\
&\quad+\frac{1}{2}\int
    \hat{\mathcal{P}}^\dagger(\mathbf{r})\;
    \bar{M}(\mathbf{r})\,\bar{\Omega}_0^2(\mathbf{r})\;
    \hat{\mathcal{P}}(\mathbf{r})
  \,d^3r.
\label{eq:H_P_dual}
\end{align}
The kinetic term has $\bar{M}^{-1}$ (not $\bar{M}$), consistent with
$p^2/2m$.  The spring-constant matrix $\bar{M}\bar{\Omega}_0^2$ has
blocks $\widetilde{M}_e\omega_{0e}^2$ and $\widetilde{M}_m\omega_{0m}^2$.

In the Power--Zienau--Woolley multipolar form, the electric coupling
is $-\int\hat{\mathbf{P}}\cdot\hat{\mathbf{E}}\,dV$ and the magnetic
coupling is $-\mu_0\int\hat{\mathbf{M}}\cdot\hat{\mathbf{H}}\,dV$.
Rewriting in dual variables and using $\mu_0/Z_0^2 = \varepsilon_0$,
both terms acquire the \emph{same} prefactor $\varepsilon_0$:
\begin{equation}
\hat{H}_{\mathcal{P}\text{-}\mathcal{E}}
= -\varepsilon_0\int
    \hat{\mathcal{P}}^\dagger(\mathbf{r})\cdot
    \hat{\mathcal{E}}(\mathbf{r})
  \,d^3r.
\label{eq:H_PE_dual}
\end{equation}
The identical coefficient $\varepsilon_0$ in both sectors is the
direct expression of electromagnetic duality invariance at the level
of the Hamiltonian: the electric and magnetic oscillators couple to
their respective field components with equal strength in the
impedance-normalized units.

Each spatial point carries a 6-component bosonic continuum, one
3-component reservoir for each sector:
\begin{equation}
\hat{H}_{\mathrm{bath}}
= \int d^3r\int_0^\infty\!d\omega\;
    \hbar\omega\;
    \hat{\mathbf{b}}^\dagger(\mathbf{r},\omega)\cdot
    \hat{\mathbf{b}}(\mathbf{r},\omega).
\label{eq:H_bath_dual}
\end{equation}

The Caldeira--Leggett coupling of each polarization sector to its
reservoir, rewritten in dual variables using the
matrix~\eqref{eq:kappa_dual}, takes the compact form
\begin{align}
\hat{H}_{\mathcal{P}\text{-bath}}
= i\int d^3r\int_0^\infty&\!d\omega\;
    \hat{\mathcal{P}}^\dagger(\mathbf{r})\;
    \bar{\kappa}(\mathbf{r},\omega)\cdot\;
    \nonumber\\ &\bigl[
      \hat{\mathbf{b}}^\dagger(\mathbf{r},\omega)
      -\hat{\mathbf{b}}(\mathbf{r},\omega)
    \bigr].
\label{eq:H_Pbath_dual}
\end{align}

%------------------------------------------------------------------
\subsection{Equations of motion and Markov limit.}
%------------------------------------------------------------------
Applying the Heisenberg equation $\dot{\hat{X}} = (i/\hbar)[\hat{H},\hat{X}]$
to $\hat{\mathcal{P}}$ and $\hat{\boldsymbol{\Pi}}_{\mathcal{P}}$ gives
the pair of first-order equations
\begin{align}
\dot{\hat{\mathcal{P}}}(\mathbf{r},t)
&= \bar{M}^{-1}(\mathbf{r})\,\hat{\boldsymbol{\Pi}}_{\mathcal{P}}(\mathbf{r},t),
\label{eq:Pdot}\\\nonumber
\dot{\hat{\boldsymbol{\Pi}}}_{\mathcal{P}}(\mathbf{r},t)
&= -\bar{M}(\mathbf{r})\,\bar{\Omega}_0^2(\mathbf{r})\,
    \hat{\mathcal{P}}(\mathbf{r},t)+\varepsilon_0\,\hat{\mathcal{E}}(\mathbf{r},t)
   \\&
   -i\int_o^\infty\,\bar{\kappa}(\mathbf{r},\omega)\,
    \bigl[\hat{\mathbf{b}}^\dagger(\mathbf r,\omega,t)-\hat{\mathbf{b}}(\mathbf r,\omega,t)\bigr].
\label{eq:Pidot}
\end{align}
Combining~\eqref{eq:Pdot} and~\eqref{eq:Pidot} by differentiating
the first and substituting the second yields the dual oscillator
equation
\begin{equation}
\ddot{\hat{\mathcal{P}}}
+\bar{\Omega}_0^2\,\hat{\mathcal{P}}
= \bar{\alpha}^{-1}\,\hat{\mathcal{E}}
  +\hat{\mathcal{F}}_N(\mathbf{r},t),
\label{eq:dual_osc}
\end{equation}
where 
$
\bar{\alpha}(\mathbf{r})
\equiv
\bar{M}(\mathbf{r})/{\varepsilon_0}$.
The dual stochastic force $\hat{\mathcal{F}}_N$ is built from the
free bath operators at $t_0$ through the back-action of
$\bar{\kappa}$ on $\hat{\mathbf{b}}$.  The Heisenberg equation for
the bath,
\begin{equation}
\dot{\hat{\mathbf{b}}}(\mathbf{r},\omega,t)
= -i\omega\,\hat{\mathbf{b}}(\mathbf{r},\omega,t)
  +\frac{\bar{\kappa}(\mathbf{r},\omega)}{\hbar}\,
   \hat{\mathcal{P}}(\mathbf{r},t),
\label{eq:bdot}
\end{equation}
is integrated formally and substituted back
into~\eqref{eq:dual_osc}.  In the broadband (Markov) limit, where
$\bar{\kappa}(\mathbf{r},\omega)$ varies slowly over the oscillator
linewidth, the retarded memory kernel reduces to instantaneous
damping and~\eqref{eq:dual_osc} becomes the damped dual Lorentz
equation
\begin{equation}
\ddot{\hat{\mathcal{P}}}
+\bar{\Gamma}\,\dot{\hat{\mathcal{P}}}
+\bar{\Omega}_0^2\,\hat{\mathcal{P}}
= \bar{\alpha}^{-1}\,\hat{\mathcal{E}}
  +\hat{\mathcal{F}}_N(\mathbf{r},t).
\label{eq:dual_lorentz}
\end{equation}
The Markov conditions that produce the damping rates
$\gamma_e(\mathbf{r})$ and $\gamma_m(\mathbf{r})$ are
\begin{align}
|\kappa_e(\mathbf{r},\omega)|^2
&= \frac{\widetilde{M}_e(\mathbf{r})\,\hbar\,\gamma_e(\mathbf{r})\,\omega}{4\pi^3},
\\ \nonumber|\kappa_m(\mathbf{r},\omega)|^2
&= \frac{\widetilde{M}_m(\mathbf{r})\,\hbar\,\gamma_m(\mathbf{r})\,\omega}{4\pi^3},
\label{eq:markov_dual}
\end{align}  
The two conditions are independent because the electric
and magnetic reservoirs never interact directly; they are fixed
separately by the requirement that each sector produces its own
Markovian damping rate upon elimination of its reservoir.

Taking the temporal Fourier transform of~\eqref{eq:dual_lorentz}
and solving for $\hat{\mathcal{P}}$ yields the frequency-domain
dual constitutive relation
\begin{equation}
\hat{\mathcal{P}}(\mathbf{r},\omega)
= \bar{\chi}(\mathbf{r},\omega)\,\hat{\mathcal{E}}(\mathbf{r},\omega)
  + \hat{\mathcal{P}}_N(\mathbf{r},\omega),
\label{eq:dual_constitutive}
\end{equation}
where the dual susceptibility tensor is
\begin{align}
\bar{\chi}(\mathbf{r},\omega)
&=\bigl(\bar{\Omega}_0^2-\omega^2-i\omega\bar{\Gamma}\bigr)^{-1}
  \bar{\alpha}^{-1}
\\ \nonumber&=
\begin{pmatrix}
  \chi_e(\mathbf{r},\omega)\,I_3 & 0 \\[4pt]
  0 & \chi_m(\mathbf{r},\omega)\,I_3
\end{pmatrix},
\label{eq:chi_dual}
\end{align}
with scalar blocks
\begin{align}
\chi_e(\mathbf{r},\omega)
&= \frac{1/\bigl(M_P(\mathbf{r})\,\varepsilon_0\bigr)}
       {\omega_{0e}^2(\mathbf{r})-\omega^2-i\gamma_e(\mathbf{r})\omega},
\\
\chi_m(\mathbf{r},\omega)
&= \frac{\mu_0/M_M(\mathbf{r})}
       {\omega_{0m}^2(\mathbf{r})-\omega^2-i\gamma_m(\mathbf{r})\omega}.
\label{eq:chi_blocks}
\end{align}
Both blocks are dimensionless, as required for a susceptibility
defined via $\hat{\mathcal{P}} = \bar{\chi}\hat{\mathcal{E}}$ between
two fields of the same dimension.
Absorbing the coherent response into the dressed dual material tensor
$\bar{\varepsilon} = I_6+\bar{\chi}$ gives the Maxwell operator
$\mathcal{M} = \mathcal{H} - k_0\bar{\varepsilon}$ of the main text.

The dual noise polarization carries the incoherent
back-action of both reservoirs,
\begin{equation}
\hat{\mathcal{P}}_N(\mathbf{r},\omega)
=\bar{\chi}_N(\mathbf{r},\omega)\,
 \hat{\mathbf{b}}(\mathbf{r},\omega,t_0),
\label{eq:PN_dual}
\end{equation}
where the noise spectrum matrix is
\begin{equation}
\bar{\chi}_N(\mathbf{r},\omega)
=
\begin{pmatrix}
  \chi_{N,e}(\mathbf{r},\omega)\,I_3 & 0 \\[4pt]
  0 & \chi_{N,m}(\mathbf{r},\omega)\,I_3
\end{pmatrix}
\label{eq:chiN_dual}
\end{equation}
with scalar amplitudes
\begin{align}
\chi_{N,e}
&= \frac{2\pi i\,\kappa_e/(M_P(\mathbf r)\varepsilon_0)}
       {\omega_{0e}^2-\omega^2-i\gamma_e\omega},
\\ 
\chi_{N,m}
&= \frac{2\pi i\,\kappa_m Z_0/M_M(\mathbf r)}
       {\omega_{0m}^2-\omega^2-i\gamma_m\omega}.
\label{eq:chiN_blocks}
\end{align}
The off-diagonal blocks of $\bar{\chi}_N$ vanish identically because
$[\hat{\mathbf{b}}_e,\hat{\mathbf{b}}_m^\dagger] = 0$
(Eq.~\eqref{eq:dualb_comm}): electric noise operators commute with
magnetic ones, so there is no cross-contamination between the two
Langevin channels.

%------------------------------------------------------------------
\subsection{Fluctuation-dissipation theorem}
%------------------------------------------------------------------
Substituting~\eqref{eq:PN_dual} into the commutator
and using the bath commutation relation~\eqref{eq:dualb_comm}, the
double spatial integral collapses to a single integral via the
delta function, and the frequency delta enforces $\omega = \omega'$.
For the electric block one obtains
$[\hat{\mathcal{P}}_{N,1,i},\hat{\mathcal{P}}_{N,1,j}^\dagger]
 = |\chi_{N,e}|^2\delta_{ij}\delta(\mathbf{r}-\mathbf{r}')\delta(\omega-\omega')$,
and the
magnetic block follows identically.  Adding both sectors and noting
that the cross-commutators vanish gives the compact six-vector form
\begin{align}
\bigl[\hat{\mathcal{P}}_{N,i}&(\mathbf{r},\omega),\,
      \hat{\mathcal{P}}_{N,j}^{\dagger}(\mathbf{r}',\omega')\bigr]
=\nonumber \\&\frac{\hbar }{\pi\epsilon_0}\,
  (\bar{\varepsilon}_I)_{ij}(\mathbf{r},\omega)\,
  \delta^{(3)}(\mathbf{r}-\mathbf{r}')\,\delta(\omega-\omega'),
\label{eq:FDT_dual}
\end{align}
where $\bar{\varepsilon}_I = (\bar{\varepsilon}-\bar{\varepsilon}^\dagger)/(2i)$
is the anti-Hermitian part of the dual material tensor from
Eq.~\eqref{eq:eps_decomp}.  This is the dual operator statement
of the fluctuation--dissipation theorem: the commutator of the
noise, which sets the minimum quantum fluctuation level in both
sectors, is proportional to the local dissipation.  Any
irreversible absorption in the medium, electric or magnetic,
necessarily injects quantum fluctuations into the corresponding
component of the dual electromagnetic field; the two effects are
inseparable consequences of the same microscopic coupling
$\bar{\kappa}$. The vanishing of the cross-commutator
$[\hat{P}_{N,i},\hat{M}_{N,j}^\dagger] = 0$ reflects the
independence of the electric and magnetic reservoirs.

With the constitutive relation~\eqref{eq:dual_constitutive}
and the noise commutator~\eqref{eq:FDT_dual} in hand,
the dressed dual Maxwell equation reads
\begin{equation}
\mathcal{M}(\mathbf{r},\omega)\,\hat{\mathcal{E}}(\mathbf{r},\omega)
= k_0\,\hat{\mathcal{P}}_N(\mathbf{r},\omega),
\label{eq:dual_Maxwell_quantum}
\end{equation}
This is a linear, driven wave equation and can be inverted with the dyadic Green's
function $\mathcal{G}$.  The general solution decomposes into a particular solution
driven by the volumetric noise source and a homogeneous solution set by the boundary
conditions on $S_{\rm in}$:
\begin{align}
\hat{\mathcal E}(\mathbf r,\omega)
&= \,k_0 \!\int\! g(\mathbf r,\mathbf r',\omega)\,
   \hat{\mathcal P}_N(\mathbf r',\omega)\,d^3 r' \nonumber\\
&\quad -i \int_{S_{\rm in}}\! g(\mathbf r,\mathbf r',\omega)\,
(\bar{n}\times)\,
\hat{\mathcal E}_{\mathrm{in}}(\mathbf r',\omega)\,d^2 r'.
\label{eq:total_field}
\end{align}
where the first term on the right corresponds to the medium assisted field and the second term is referred to as the boundary-assisted contribution.

%% file: references.bib
@article{robicheaux2021beyond,
	title        = {Beyond Lowest Order Mean-Field Theory for Light Interacting with Atom Arrays},
	author       = {Robicheaux, Francis and Suresh, Deepak A.},
	year         = 2021,
	journal      = {Phys. Rev. A},
	volume       = 104,
	pages        = {023702},
	doi          = {10.1103/PhysRevA.104.023702},
	url          = {https://doi.org/10.1103/PhysRevA.104.023702}
}

@article{scheel2009duality,
  title        = {Macroscopic Quantum Electrodynamics and Duality},
  author       = {Scheel, Stefan and Buhmann, Stefan Yoshi},
  journal      = {Phys. Rev. Lett.},
  volume       = {102},
  pages        = {140404},
  year         = {2009},
  doi          = {10.1103/PhysRevLett.102.140404},
  url          = {https://doi.org/10.1103/PhysRevLett.102.140404}
}

@article{scheel2012macroscopic,
  title        = {Macroscopic Quantum Electrodynamics in Nonlocal and Nonreciprocal Media},
  author       = {Scheel, Stefan and Buhmann, Stefan Yoshi},
  journal      = {New J. Phys.},
  volume       = {14},
  pages        = {083034},
  year         = {2012},
  doi          = {10.1088/1367-2630/14/8/083034},
  url          = {https://doi.org/10.1088/1367-2630/14/8/083034}
}

@article{rapp2025purcell,
  title        = {Purcell Effect in Chiral Environments: A Macroscopic {QED} Perspective},
  author       = {Rapp, C. S. and Franz, Janine C. and Buhmann, Stefan Yoshi and Franca, O. J.},
  journal      = {Phys. Rev. A},
  volume       = {112},
  pages        = {013109},
  year         = {2025},
  doi          = {10.1103/PhysRevA.112.013109},
  url          = {https://doi.org/10.1103/PhysRevA.112.013109}
}

@article{ThomasWeiss2019FirstOrderPert,
author = {S. Both and T. Weiss},
journal = {Opt. Lett.},
number = {24},
pages = {5917--5920},
publisher = {Optica Publishing Group},
title = {First-order perturbation theory for changes in the surrounding of open optical resonators},
volume = {44},
month = {Dec},
year = {2019},
url = {https://opg.optica.org/ol/abstract.cfm?URI=ol-44-24-5917},
doi = {10.1364/OL.44.005917},
}

@article{lodahl2015interfacing,
	title        = {Interfacing Single Photons and Single Quantum Dots with Photonic Nanostructures},
	author       = {Lodahl, Peter and Mahmoodian, Sahand and Stobbe, S{\"o}ren},
	year         = 2015,
	journal      = {Rev. Mod. Phys.},
	volume       = 87,
	number       = 2,
	pages        = {347--400},
	doi          = {10.1103/RevModPhys.87.347},
	url          = {https://doi.org/10.1103/RevModPhys.87.347}
}

@article{masson2022universality,
	title        = {Universality of Dicke superradiance in arrays of quantum emitters},
	author       = {Masson, Stuart J. and Asenjo-Garcia, Ana},
	year         = 2022,
	journal      = {Nat. Commun.},
	volume       = 13,
	pages        = 2285,
	doi          = {10.1038/s41467-022-29805-4},
	url          = {https://doi.org/10.1038/s41467-022-29805-4}
}

@article{masson2024multilevelatom,
  title = {Dicke Superradiance in Ordered Arrays of Multilevel Atoms},
  author = {Masson, Stuart J. and Covey, Jacob P. and Will, Sebastian and Asenjo-Garcia, Ana},
  journal = {PRX Quantum},
  volume = {5},
  issue = {1},
  pages = {010344},
  numpages = {19},
  year = {2024},
  month = {Mar},
  publisher = {American Physical Society},
  doi = {10.1103/PRXQuantum.5.010344},
  url = {https://link.aps.org/doi/10.1103/PRXQuantum.5.010344}
}

@article{asenjo-garcia2017exponential,
	title        = {Exponential Improvement in Photon Storage Fidelities Using Subradiance and ``Selective Radiance'' in Atomic Arrays},
	author       = {Asenjo-Garcia, Ana and Moreno-Cardoner, Manuel and Albrecht, Andreas and Kimble, H. J. and Chang, Darrick E.},
	year         = 2017,
	journal      = {Phys. Rev. X},
	volume       = 7,
	pages        = {031024},
	doi          = {10.1103/PhysRevX.7.031024},
	url          = {https://doi.org/10.1103/PhysRevX.7.031024}
}

@article{martin-cano2011entanglement,
	title        = {Entanglement of two qubits mediated by one-dimensional plasmonic waveguides},
	author       = {Mart{\'\i}n-Cano, Diego and Gonz{\'a}lez-Tudela, Alejandro and Mart{\'\i}n-Moreno, Luis and Garc{\'\i}a-Vidal, F. J. and Tejedor, Carlos and Moreno, Enrique},
	year         = 2011,
	journal      = {Phys. Rev. B},
	volume       = 84,
	pages        = 235306,
	doi          = {10.1103/PhysRevB.84.235306},
	url          = {https://doi.org/10.1103/PhysRevB.84.235306}
}

@article{shen2005coherent,
	title        = {Coherent single photon transport in a one-dimensional waveguide coupled with superconducting quantum bits},
	author       = {Shen, Jung-Tsung and Fan, Shanhui},
	year         = 2005,
	journal      = {Phys. Rev. Lett.},
	volume       = 95,
	pages        = 213001,
	doi          = {10.1103/PhysRevLett.95.213001},
	url          = {https://doi.org/10.1103/PhysRevLett.95.213001}
}

@book{scully1997quantum,
	title        = {Quantum Optics},
	author       = {Scully, Marlan O. and Zubairy, M. Suhail},
	year         = 1997,
	publisher    = {Cambridge Univ. Press},
	address      = {Cambridge},
	doi          = {10.1017/CBO9780511813993},
	url          = {https://doi.org/10.1017/CBO9780511813993}
}

@article{kimble1998strong,
	title        = {Strong interactions of single atoms and photons in cavity {QED}},
	author       = {Kimble, H. J.},
	year         = 1998,
	journal      = {Phys. Scr.},
	volume       = {T76},
	pages        = {127--137},
	doi          = {10.1238/Physica.Topical.076a00127},
	url          = {https://doi.org/10.1238/Physica.Topical.076a00127}
}

@article{glauber1963quantum,
	title        = {The Quantum Theory of Optical Coherence},
	author       = {Glauber, Roy J.},
	year         = 1963,
	journal      = {Phys. Rev.},
	publisher    = {American Physical Society},
	volume       = 130,
	number       = 6,
	pages        = {2529--2539},
	doi          = {10.1103/PhysRev.130.2529},
	url          = {https://doi.org/10.1103/PhysRev.130.2529}
}

@article{raimond2001manipulating,
	title        = {Manipulating quantum entanglement with atoms and photons in a cavity},
	author       = {Raimond, Jean-Michel and Brune, Michel and Haroche, Serge},
	year         = 2001,
	journal      = {Rev. Mod. Phys.},
	publisher    = {American Physical Society},
	volume       = 73,
	number       = 3,
	pages        = {565--582},
	doi          = {10.1103/RevModPhys.73.565},
	url          = {https://doi.org/10.1103/RevModPhys.73.565}
}

@book{collin1990field,
	title        = {Field Theory of Guided Waves},
	author       = {Collin, Robert E.},
	year         = 1990,
	publisher    = {Wiley-IEEE Press},
	address      = {New York, NY},
	series       = {IEEE Press Series on Electromagnetic Wave Theory},
	isbn         = 9780879422370,
	edition      = 2
}

@book{chew1995waves1,
	title        = {Waves and Fields in Inhomogeneous Media},
	author       = {Chew, Weng Cho},
	year         = 1995,
	publisher    = {IEEE Press},
	address      = {Piscataway, NJ},
	series       = {IEEE Press Series on Electromagnetic Waves},
	isbn         = 9780780311169
}

@book{hanson2002operator,
  author    = {Hanson, George W. and Yakovlev, Alexander B.},
  title     = {Operator Theory for Electromagnetics: An Introduction},
  publisher = {Springer},
  address   = {New York, NY},
  year      = {2002},
  isbn      = {978-0-387-95278-9},
  doi       = {10.1007/978-1-4757-3679-3},
}

@book{dudley1994mathematical,
	title        = {Mathematical Foundations for Electromagnetic Theory},
	author       = {Dudley, Donald G.},
	year         = 1994,
	publisher    = {Wiley-IEEE Press},
	address      = {Piscataway, NJ},
	isbn         = 9780780310223
}

@book{tai1994dyadic1,
  author    = {Tai, Chen-To},
  title     = {Dyadic Green Functions in Electromagnetic Theory},
  edition   = {2},
  publisher = {IEEE Press},
  address   = {Piscataway, NJ},
  year      = {1994},
  isbn      = {9780780304499},
}

@article{collin1986dyadic,
	title        = {The dyadic Green's function as an inverse operator},
	author       = {Collin, R. E.},
	year         = 1986,
	journal      = {Radio Sci.},
	volume       = 21,
	number       = 6,
	pages        = {883--890},
	doi          = {10.1029/RS021i006p00883}
}

@article{lange2024superradiant,
  title        = {Superradiant and Subradiant States in Lifetime-Limited Organic Molecules through Laser-Induced Tuning},
  author       = {Lange, Christian M. and Daggett, Emma and Walther, Valentin and Huang, Libai and Hood, Jonathan D.},
  journal      = {Nat. Phys.},
  volume       = {20},
  number       = {5},
  pages        = {836--842},
  year         = {2024},
  doi          = {10.1038/s41567-024-02404-4},
  url          = {https://doi.org/10.1038/s41567-024-02404-4}
}

@article{lange2026hybrid,
  title        = {A Hybrid Molecular--Nanophotonic Platform for On-Chip Cavity Quantum Electrodynamics and Collective Interactions},
  author       = {Lange, Christian M. and Keni, Arya D. and Agarwal, Ishita and Daggett, Emma and Mansukhani, Adhyyan S. and Kundu, Ankit and Cerjan, Benjamin and Huang, Libai and Hood, Jonathan D.},
  journal      = {ACS Nano},
  year         = {2026},
  doi          = {10.1021/acsnano.5c19465},
  url          = {https://doi.org/10.1021/acsnano.5c19465}
  }

@article{franke2020fluctuation,
  title        = {Fluctuation-Dissipation Theorem and Fundamental Photon Commutation Relations in Lossy Nanostructures Using Quasinormal Modes},
  author       = {Franke, Sebastian and Ren, Juanjuan and Hughes, Stephen and Richter, Marten},
  journal      = {Phys. Rev. Research},
  volume       = {2},
  pages        = {033332},
  year         = {2020},
  doi          = {10.1103/PhysRevResearch.2.033332},
  url          = {https://doi.org/10.1103/PhysRevResearch.2.033332}
}

@article{kristensen2014quasinormal,
  title        = {Quasinormal Mode Approach to Modelling Light-Emission and Propagation in Nanoplasmonics},
  author       = {Kristensen, Peter T. and Young, Jeff F. and Hughes, Stephen},
  journal      = {New J. Phys.},
  volume       = {16},
  pages        = {113048},
  year         = {2014},
  doi          = {10.1088/1367-2630/16/11/113048},
  url          = {https://doi.org/10.1088/1367-2630/16/11/113048}
}

@article{hood2016atom,
  title        = {Atom--atom interactions around the band edge of a photonic crystal waveguide},
  author       = {Hood, Jonathan D. and Goban, Akihisa and Asenjo-Garcia, Ana and Lu, Mingwu and Yu, Su-Peng and Chang, Darrick E. and Kimble, H. J.},
  journal      = {Proc. Natl. Acad. Sci. U.S.A.},
  volume       = {113},
  number       = {38},
  pages        = {10507--10512},
  year         = {2016},
  doi          = {10.1073/pnas.1603788113},
  url          = {https://doi.org/10.1073/pnas.1603788113}
}

@article{newton1976optical,
	title        = {Optical theorem and beyond},
	author       = {Newton, Roger G.},
	year         = 1976,
	journal      = {Am. J. Phys.},
	volume       = 44,
	number       = 7,
	pages        = {639--642},
	doi          = {10.1119/1.10324}
}

@article{asenjo-garcia2017atom,
  title        = {Atom-light interactions in quasi-one-dimensional nanostructures: A Green's-function perspective},
  author       = {Asenjo-Garcia, Ana and Hood, Jonathan D. and Chang, Darrick E. and Kimble, H. J.},
  journal      = {Phys. Rev. A},
  volume       = {95},
  pages        = {033818},
  year         = {2017},
  doi          = {10.1103/PhysRevA.95.033818},
  url          = {https://doi.org/10.1103/PhysRevA.95.033818}
}

@book{newton1982scattering,
	title        = {Scattering Theory of Waves and Particles},
	author       = {Newton, Roger G.},
	year         = 1982,
	publisher    = {Springer-Verlag},
	address      = {New York, NY},
	doi          = {10.1007/978-3-642-88128-2},
	edition      = 2
}

@article{johnson2002adiabatic,
	title        = {Adiabatic theorem and continuous coupled-mode theory for efficient taper transitions in photonic crystals},
	author       = {Johnson, Steven G. and Bienstman, Peter and Skorobogatiy, M. A. and Ibanescu, Mihai and Lidorikis, Elefterios and Joannopoulos, J. D.},
	year         = 2002,
	month        = {Dec},
	journal      = {Phys. Rev. E},
	publisher    = {American Physical Society},
	volume       = 66,
	pages        = {066608},
	doi          = {10.1103/PhysRevE.66.066608},
	url          = {https://link.aps.org/doi/10.1103/PhysRevE.66.066608},
	issue        = 6,
	numpages     = 15
}

@article{buhmann2007dispersion,
	title        = {Dispersion {{Forces}} in {{Macroscopic}} Quantum Electrodynamics},
	author       = {Buhmann, Stefan Yoshi and Welsch, Dirk-Gunnar},
	year         = 2007,
	journal      = {Prog. Quantum Electron.},
	volume       = 31,
	number       = 2,
	pages        = {51--130},
	doi          = {10.1016/j.pquantelec.2007.03.001},
	issn         = {0079-6727},
	url          = {https://www.sciencedirect.com/science/article/pii/S0079672707000249},
	shortjournal = {Prog. Quantum Electron.},
	eprint       = {quant-ph/0608118v2},
	eprinttype   = {arXiv}
}

@article{ciattoni20240719quantum,
	title        = {Quantum Electrodynamics of Lossy Magnetodielectric Samples in Vacuum: {{Modified Langevin}} Noise Formalism},
	shorttitle   = {Quantum Electrodynamics of Lossy Magnetodielectric Samples in Vacuum},
	author       = {Ciattoni, A.},
	year         = {2024},
	journal      = {Phys. Rev. A},
	publisher    = {American Physical Society},
	volume       = 110,
	number       = 1,
	pages        = {013707},
	doi          = {10.1103/PhysRevA.110.013707},
	issn         = {2469-9926, 2469-9934},
	url          = {https://link.aps.org/doi/10.1103/PhysRevA.110.013707},
	shortjournal = {Phys. Rev. A}
}

@article{drezet2017equivalence,
	title        = {Equivalence between the {{Hamiltonian}} and {{Langevin}} Noise Descriptions of Plasmon Polaritons in a Dispersive and Lossy Inhomogeneous Medium},
	author       = {Drezet, Aurélien},
	year         = {2017},
	journal      = {Phys. Rev. A},
	volume       = 96,
	number       = 3,
	pages        = {033849},
	doi          = {10.1103/PhysRevA.96.033849},
	issn         = {2469-9926, 2469-9934},
	url          = {https://link.aps.org/doi/10.1103/PhysRevA.96.033849},
	shortjournal = {Phys. Rev. A}
}

@article{drezet2017quantizing,
	title        = {Quantizing Polaritons in Inhomogeneous Dissipative Systems},
	author       = {Drezet, Aurélien},
	journal      = {Phys. Rev. A},
    year         = {2017},
	volume       = 95,
	number       = 2,
	pages        = {023831},
	doi          = {10.1103/PhysRevA.95.023831},
	issn         = {2469-9926, 2469-9934},
	url          = {https://link.aps.org/doi/10.1103/PhysRevA.95.023831},
	urldate      = {2025-06-17},
	date         = {2017-02-16}
}

@article{gruner1996green-function,
	title        = {Green-Function Approach to the Radiation-Field Quantization for Homogeneous and Inhomogeneous {{Kramers-Kronig}} Dielectrics},
	author       = {Gruner, T. and Welsch, D. G.},
	year         = 1996,
	journal      = {Phys. Rev. A},
	publisher    = {American Physical Society},
	volume       = 53,
	number       = 3,
	pages        = {1818--1829},
	doi          = {10.1103/PhysRevA.53.1818},
	issn         = 10941622,
	url          = {https://link.aps.org/doi/10.1103/PhysRevA.53.1818},
	shortjournal = {Phys. Rev. A}
}

@article{huttner1992quantization,
	title        = {Quantization of the Electromagnetic Field in Dielectrics},
	author       = {Huttner, Bruno and Barnett, Stephen M},
	year         = 1992,
	journal      = {Phys. Rev. A},
	publisher    = {American Physical Society},
	volume       = 46,
	number       = 7,
	pages        = {4306--4322},
	doi          = {10.1103/PhysRevA.46.4306},
	url          = {http://pra.aps.org/abstract/PRA/v46/i7/p4306_1},
	shortjournal = {Phys. Rev. A}
}

@article{scheel2006quantum,
  title = {Quantum Theory of Light and Noise Polarization in Nonlinear Optics},
  author = {Scheel, Stefan and Welsch, Dirk-Gunnar},
  journal = {Phys. Rev. Lett.},
  volume = {96},
  issue = {7},
  pages = {073601},
  numpages = {4},
  year = {2006},
  month = {Feb},
  publisher = {American Physical Society},
  doi = {10.1103/PhysRevLett.96.073601},
  url = {https://link.aps.org/doi/10.1103/PhysRevLett.96.073601}
}

@article{molesky2018inverse,
	title        = {Inverse design in nanophotonics},
	author       = {Molesky, Sean and Lin, Zin and Piggott, Alexander Y. and Jin, Weiliang and Vuckovic, Jelena and Rodriguez, Alejandro W.},
	year         = 2018,
	month        = nov,
	journal      = {Nat. Photonics},
	publisher    = {Springer Nature},
	volume       = 12,
	number       = 11,
	pages        = {659--670},
	doi          = {10.1038/s41566-018-0246-9}
}

@article{yelin2022superradiance,
	title        = {Superradiance and subradiance in inverted atomic arrays},
	author       = {Rubies-Bigorda, Oriol and Yelin, Susanne F.},
	year         = 2022,
	month        = {Nov},
	journal      = {Phys. Rev. A},
	publisher    = {American Physical Society},
	volume       = 106,
	pages        = {053717},
	doi          = {10.1103/PhysRevA.106.053717},
	url          = {https://link.aps.org/doi/10.1103/PhysRevA.106.053717},
	issue        = 5,
	numpages     = 12
}

@article{stefano2001mode,
	title        = {Mode Expansion and Photon Operators in Dispersive and Absorbing Dielectrics},
	author       = {Stefano, Omar Di and , Salvatore, Savasta},
	year         = {2001},
	journal      = {J. Mod. Opt.},
	publisher    = {Taylor and Francis},
	volume       = 48,
	number       = 1,
	pages        = {67--84},
	doi          = {10.1080/09500340108235155},
	issn         = {0950-0340},
	url          = {https://doi.org/10.1080/09500340108235155},
	shortjournal = {J. Mod. Opt.}
}

@article{glauber1991quantum,
	title        = {Quantum optics of dielectric media},
	author       = {Glauber, Roy J. and Lewenstein, M.},
	year         = 1991,
	month        = {Jan},
	journal      = {Phys. Rev. A},
	publisher    = {American Physical Society},
	volume       = 43,
	pages        = {467--491},
	doi          = {10.1103/PhysRevA.43.467},
	url          = {https://link.aps.org/doi/10.1103/PhysRevA.43.467},
	issue        = 1,
	numpages     = {0}
}

@article{knoll1987action,
	title        = {Action of passive, lossless optical systems in quantum optics},
	author       = {Kn\"oll, L. and Vogel, W. and Welsch, D. -G.},
	year         = 1987,
	month        = {Oct},
	journal      = {Phys. Rev. A},
	publisher    = {American Physical Society},
	volume       = 36,
	pages        = {3803--3818},
	doi          = {10.1103/PhysRevA.36.3803},
	url          = {https://link.aps.org/doi/10.1103/PhysRevA.36.3803},
	issue        = 8,
	numpages     = {0}
}

@article{berreman1972optics,
	title        = {Optics in Stratified and Anisotropic Media: 4{\texttimes}4-Matrix Formulation},
	author       = {Dwight W. Berreman},
	year         = 1972,
	month        = {Apr},
	journal      = {J. Opt. Soc. Am.},
	publisher    = {Optica Publishing Group},
	volume       = 62,
	number       = 4,
	pages        = {502--510},
	doi          = {10.1364/JOSA.62.000502},
	url          = {https://opg.optica.org/abstract.cfm?URI=josa-62-4-502}
}

@article{ciattoni2025quantum-optical,
  title = {Quantum-optical scattering by macroscopic lossy objects: A general approach},
  author = {Ciattoni, A.},
  journal = {Phys. Rev. A},
  volume = {112},
  issue = {1},
  pages = {013704},
  numpages = {37},
  year = {2025},
  month = {Jul},
  publisher = {American Physical Society},
  doi = {10.1103/9mpv-thdp},
  url = {https://link.aps.org/doi/10.1103/9mpv-thdp}
}

@article{sheremet2023waveguide,
	title        = {Waveguide quantum electrodynamics: Collective radiance and photon-photon correlations},
	author       = {Sheremet, Alexandra S. and Petrov, Mihail I. and Iorsh, Ivan V. and Poshakinskiy, Alexander V. and Poddubny, Alexander N.},
	year         = 2023,
	month        = {Mar},
	journal      = {Rev. Mod. Phys.},
	publisher    = {American Physical Society},
	volume       = 95,
	pages        = {015002},
	doi          = {10.1103/RevModPhys.95.015002},
	url          = {https://link.aps.org/doi/10.1103/RevModPhys.95.015002},
	issue        = 1,
	numpages     = 59
}

@article{schelkunoff1936some,
	title        = {Some equivalence theorems of electromagnetics and their application to radiation problems},
	author       = {Schelkunoff, S. A.},
	year         = 1936,
	journal      = {Bell Syst. Tech. J.},
	volume       = 15,
	number       = 1,
	pages        = {92--112},
	doi          = {10.1002/j.1538-7305.1936.tb00720.x}
}

@book{novotny2006principles,
	title        = {Principles of Nano-Optics},
	author       = {Novotny, Lukas and Hecht, Bert},
	year         = 2006,
	publisher    = {Cambridge University Press},
	place        = {Cambridge}
}

@book{yoshibuhmann2013dispersion,
	title        = {Dispersion Forces I: Macroscopic Quantum Electrodynamics and Ground-State Casimir, Casimir–Polder and van der Waals Forces},
	author       = {Stefan Yoshi Buhmann},
	year         = 2013,
	publisher    = {Springer Berlin, Heidelberg},
	isbn         = {978-3-642-32484-0}
}

@article{Yaghjian1980,
	title        = {Electric dyadic Green's functions in the source region},
	author       = {Yaghjian, A.D.},
	year         = 1980,
	journal      = {Proc. IEEE},
	volume       = 68,
	number       = 2,
	pages        = {248--263},
	doi          = {10.1109/PROC.1980.11620}
}

@misc{rosa2010quantum,
      title={Quantum Fields in a Dielectric: Langevin and Exact Diagonalization Approaches}, 
      author={F. S. S. Rosa and D. A. R. Dalvit and P. W. Milonni},
      year={2009},
      eprint={0912.0279},
      archivePrefix={arXiv},
      primaryClass={quant-ph},
      url={https://arxiv.org/abs/0912.0279}, 
}

@article{silveirinha2018topological,
  title = {Topological classification of Chern-type insulators by means of the photonic Green function},
  author = {Silveirinha, M\'ario G.},
  journal = {Phys. Rev. B},
  volume = {97},
  issue = {11},
  pages = {115146},
  numpages = {15},
  year = {2018},
  month = {Mar},
  publisher = {American Physical Society},
  doi = {10.1103/PhysRevB.97.115146},
  url = {https://link.aps.org/doi/10.1103/PhysRevB.97.115146}
}

@incollection{silveirinha2021modal,
  author    = {Silveirinha, M{\'a}rio G.},
  title     = {Modal Expansions in Dispersive Material Systems with Application to Quantum Optics and Topological Photonics},
  booktitle = {Advances in Mathematical Methods for Electromagnetics},
  editor    = {Smith, Paul and Kobayashi, Kazuya},
  publisher = {IET},
  year      = {2020},
  chapter   = {24},
  doi       = {10.1049/SBEW528E\_ch24},
}

@article{Robicheaux2020directional,
  title = {Theoretical study of early-time superradiance for atom clouds and arrays},
  author = {Robicheaux, F.},
  journal = {Phys. Rev. A},
  volume = {104},
  issue = {6},
  pages = {063706},
  numpages = {10},
  year = {2021},
  month = {Dec},
  publisher = {American Physical Society},
  doi = {10.1103/PhysRevA.104.063706},
  url = {https://link.aps.org/doi/10.1103/PhysRevA.104.063706}
}

@article{woolley2020power,
  title = {Power-Zienau-Woolley representations of nonrelativistic {QED} for atoms and molecules},
  author = {Woolley, R. Guy},
  journal = {Phys. Rev. Res.},
  volume = {2},
  issue = {1},
  pages = {013206},
  numpages = {11},
  year = {2020},
  month = {Feb},
  publisher = {American Physical Society},
  doi = {10.1103/PhysRevResearch.2.013206},
  url = {https://link.aps.org/doi/10.1103/PhysRevResearch.2.013206}
}

@article{power_zienau,
 author = {Power, E. A. and Zienau, S.},
 ISSN = {00804614},
 URL = {http://www.jstor.org/stable/91664},
 journal = {Philosophical Transactions of the Royal Society of London. Series A, Mathematical and Physical Sciences},
 number = {999},
 pages = {427--454},
 publisher = {Royal Society},
 title = {Coulomb Gauge in Non-Relativistic Quantum Electro-Dynamics and the Shape of Spectral Lines},
 urldate = {2026-03-26},
 volume = {251},
 year = {1959}
}

@book{landau1984electrodynamics,
  author    = {L. D. Landau and E. M. Lifshitz and L. P. Pitaevskii},
  title     = {Electrodynamics of Continuous Media},
  series    = {Course of Theoretical Physics},
  volume    = {8},
  edition   = {2},
  publisher = {Pergamon Press},
  address   = {Oxford},
  year      = {1984},
  isbn      = {978-0-08-030275-1}
}

@article{rosa2010electromagnetic,
  author  = {F. S. S. Rosa and D. A. R. Dalvit and P. W. Milonni},
  title   = {Electromagnetic Energy, Absorption, and Casimir Forces. I. Uniform Dielectric Media in Thermal Equilibrium},
  journal = {Physical Review A},
  volume  = {81},
  number  = {3},
  pages   = {033812},
  year    = {2010},
  doi     = {10.1103/PhysRevA.81.033812}
}
